\documentclass[journal,twoside,web]{ieeecolor}
\usepackage{booktabs}
\usepackage{multirow}
\usepackage{array}
\usepackage{tmi}
\usepackage{cite}
\usepackage{amsmath,amssymb,amsfonts}
\usepackage{mathrsfs} 
\usepackage{algorithmic}
\usepackage{graphicx}
\usepackage{textcomp}
\usepackage{comment}
\usepackage{pifont}
\usepackage{upgreek}

\usepackage{hyperref}
\hypersetup{
    colorlinks=true,
    linkcolor=red,  
    urlcolor=blue,
    citecolor=red,
    }

\def\BibTeX{{\rm B\kern-.05em{\sc i\kern-.025em b}\kern-.08em
    T\kern-.1667em\lower.7ex\hbox{E}\kern-.125emX}}
\newcommand{\black}[1]{\textcolor{black}{#1}}
\newcommand{\blue}[1]{\textcolor{black}{#1}}

\begin{document}

\title{Chest X-ray Foundation Model with Global and Local Representations Integration}
% \title{CheXFound: Advancing Chest X-ray Foundation Model for Robust Disease Detection with Global and Local Representations Integration}

\author{Zefan Yang, Xuanang Xu, Jiajin Zhang, Ge Wang, \IEEEmembership{Fellow, IEEE}, \\Mannudeep K. Kalra, and Pingkun Yan, \IEEEmembership{Senior Member, IEEE}
\thanks{Z. Yang, X. Xu, G. Wang, and P. Yan* are with the Department of Biomedical Engineering and Center for Biotechnology and Interdisciplinary Studies, Rensselaer Polytechnic Institute, Troy, NY, USA.}
\thanks{J. Zhang was with the Department of Biomedical Engineering and Center for Biotechnology and Interdisciplinary Studies, Rensselaer Polytechnic Institute, when he contributed to this work.}
\thanks{M. K. Kalra is with the Department of Radiology, Massachusetts General Hospital, Harvard Medical School, Boston, MA, USA.}
\thanks{This research was funded in part by  the NSF CAREER award 2046708 and the NIA Predoctoral Training Program For Alzheimer’s Disease At The Interface Of Data Science, Engineering And Biology Training Program T32AG078123.}
\thanks{*Corresponding author: yanp2@rpi.edu}
}
% \py{Please get the official name of the training grant.}

\maketitle

\begin{abstract}
% CXR (CXR) is one of the most commonly-ordered imaging test worldwide.
% Clinical CXR interpretation involves a wide range of tasks, from detecting thoracic diseases to monitoring postoperative recovery.
% With the vast array of tasks, training specialized classification models for each task has limitations. 
% These models are only effective for limited pathologies and have inferior generalizability to out-of-distribution datasets.
% They also requires a large collection of cost-prohibitive labels to learn meaningful representations.
% To address these limitations, we introduce CheXFound, a self-supervised vision foundation model containing robust CXR representations that can generalize to a wide range of downstream tasks.
Chest X-ray (CXR) is the most frequently ordered imaging test, supporting diverse clinical tasks from thoracic disease detection to postoperative monitoring. However, task-specific classification models are limited in scope, require costly labeled data, and lack generalizability to out-of-distribution datasets. To address these challenges, we introduce CheXFound, a self-supervised vision foundation model that learns robust CXR representations and generalizes effectively across a wide range of downstream tasks.
We pretrained CheXFound on a curated CXR-987K dataset, comprising over approximately 987K unique CXRs from 12 publicly available sources. We propose a Global and Local Representations Integration (GLoRI) head for downstream adaptations, by incorporating fine- and coarse-grained disease-specific local features with global image features for enhanced performance in multilabel classification.
Our experimental results showed that CheXFound outperformed state-of-the-art models in classifying 40 disease findings across different prevalence levels on the CXR-LT 24 dataset and exhibited superior label efficiency on downstream tasks with limited training data.
Additionally, CheXFound achieved significant improvements on downstream tasks with out-of-distribution datasets, including opportunistic cardiovascular disease risk estimation, mortality prediction, malpositioned tube detection, and anatomical structure segmentation.
The above results demonstrate CheXFound's strong generalization capabilities, which will enable diverse downstream adaptations with improved label efficiency in future applications. The project source code is publicly available at \url{https://github.com/RPIDIAL/CheXFound}.
\end{abstract}


\section{Introduction}
\label{sec:introduction}

\IEEEPARstart{C}{hest} X-ray (CXR) is one of the most commonly-ordered imaging tests worldwide \cite{raoof2012interpretation}.
Clinical CXR interpretation encompasses a broad spectrum of tasks, including detecting diseases associated with the lungs, heart, blood vessel, and bones, as well as monitoring postoperative recovery and the positioning of support devices. 
With advancements in computer-aided diagnosis, these tasks now extend even further to include opportunistic disease risk assessment, such as cardiovascular disease\cite{Weiss24-CVD-CXR,yang2024cardiovascular}, mortality risk\cite{weiss2023deep}, and diabetes\cite{pyrros2023opportunistic}, among other factors not directly quantifiable by human eyes.
% \cite{kamel2021prediction, ueda2023artificial}
Training specialized classification models for each task from scratch poses significant limitations. 
Such models are typically effective only within a narrow scope of pathologies and struggle to generalize to out-of-distribution datasets.
Furthermore, developing these models \blue{requires} extensive labeled datasets, which are both cost-prohibitive and inefficient. These challenges underscore the need for self-supervised models that can learn robust representations and demonstrate superior generalization capabilities across diverse tasks.
% since it is cost prohibitive to have radiologists annotate a large\ins{-}scale dataset.
% Small-scale, task-specific training can only generalize to a limited scope of disease findings and has performance discrepancies between development and deployment due to data distribution shifts.

Recent advancements in the field of computer vision\cite{chen2021empirical, he2022masked, caron2021emerging, zhou2021ibot} demonstrate that self-supervised vision models can produce task-agnostic and semantic-rich image representations that achieve improved performance on a broad spectrum of downstream tasks.
% and computational pathology\cite{chen2024towards, wang2024pathology}
% \rev{general-purpose}
% transcend previous models
% predominantly
Such models are called foundation models because of their superior capabilities to adapt to diverse downstream tasks when pretrained on large-scale data. 
% the large-scale pretraining and 
% they are pretrained with large-scale data
% This generalization capability is heavily reliant on the massive size and diversity of pretraining data.
Recent works in self-supervised learning for CXR interpretation adopt a series of advanced training strategies to learn high-quality image representations, including contrastive learning \cite{azizi2023robust}, masked image modeling (MIM) \cite{yao2024eva}, and self-distillation \cite{perez2024rad}. Research further use CXRs and their clinical reports to perform contrastive language-image pretraining \cite{tiu2022expert}.
However, these studies have two major limitations. 
\blue{First, most of these proposed models are constrained by either model capacity or pretraining data size, limiting their capability to learn rich and effective representations for long-tail classification, as demonstrated by our detailed experimental results in Fig. \ref{fig:radar_cxrlt}, showing that existing foundation models achieved degraded performance across all 40 diseases compared to CheXFound with large model capacity and data size.
}
% For example, EVA-X \cite{yao2024eva} and RAD-DINO \cite{perez2024rad} used the ViT-Base architecture with only 86M parameters and were pretrained on up to five public data sources.
% First, they only evaluate model performance for classifying a narrow range of disease findings, without considering the long-tail nature of pathologies in CXR and the opportunistic CXR interpretation tasks, such as cardiovascular disease (CVD) risk estimation and mortality prediction.
Second, most of these studies simply rely on the global image features for disease detection, overlooking the use of patch embeddings learned by the foundation model to provide disease-specific local features to enhance performance.
% reduce ambiguities.
% arisen from merely using the global representation.
% in multilabel classification.
% the model capabilities for 
% Second, previous studies focused on detecting disease findings recorded in radiology reports and did not evaluate the generalization capability of the learned representations on tasks that extend radiologists ability, including cardiovascular disease risk, coronary calcium scores, ejection fraction, all-cause mortality, etc. 
Addressing these limitations is pivotal to the development of the CXR foundation models towards clinical applications which often involve interpreting a wide range of disease findings. It also has broader implications by enabling CVD risk estimation and mortality prediction with a routine CXR.
% in risk estimation of diseases that radiologists \rev{cannot}{may not} see.

In this work, we introduce CheXFound, a vision foundation model specialized for CXR image analysis that learns high-quality CXR representations and generalizes effectively across a wide range of thoracic disease detection and opportunistic risk estimation tasks. 
% To pretrain our CheXFound model, we curate CXR-1M, a large-scale CXR dataset containing
% more than one million
We pretrained CheXFound on a curated \blue{CXR-987K} dataset, comprising \blue{approximately 987K} unique CXRs from \blue{12} publicly available datasets, including MIMIC-CXR \cite{johnson2019mimic}, CheXpert \cite{irvin2019chexpert}, PadChest \cite{bustos2020padchest}, NIH-CXR \cite{wang2017chestx}, BRAX \cite{reis2022brax}, and CANDID-PTX \cite{feng2021curation}, among others.
Our CheXFound model was pretrained via DINOv2 \cite{oquab2023dinov2}, \blue{with specific adaptations in the model initialization strategy and pretraining objective weight to enhance its effectiveness in the CXR domain.}
% , a state-of-the-art self-supervised learning method with strong off-the-shelf linear probe performance.
% , VinDr-CXR \cite{nguyen2022vindr}
 % released by a range of institutes
% \ins{Our CheXFound model is} pretrained \del{using the vision transformer architecture (ViT-Large)} on a large cohort of \ins{[one million]} CXRs from \rev{multiple}{[13]} institutions \com{Explicitly show the numbers of CXRs and institutes involved in this study to impress the reviewers. BTW, I think our data curation (CXR-1M) is also a major contribution in this work. We can emphasize it here.} \rev{. In the pretraining stage, we use}{via} a self-supervised learning method named DINOv2\cite{oquab2023dinov2}, which uses self-distillation techniques achieving strong image representations for downstream prediction tasks.
% most of the time, include more information to construct a complete sentence.
For downstream adaptation, we introduce a Global and Local Representations Integration (GLoRI) head 
that was trained on top of the frozen CheXFound model \blue{to perform attention probing}. 
GLoRI employs the attention mechanism to compute \blue{fine- and coarse-grained} disease-specific local features, which are then integrated with the global image features to improve multilabel classification performance.
 % from the \texttt{[CLS]} token
 % in CXR interpretation
% GLoRI computes attention-pooled CXR representations for each disease finding and integrates the global features from the [CLS] token to address the multilabel classification problem.
% \py{What is after pretraining? Should also mention Glori here.}
% Contributions on data curation

% two tiers
We assessed CheXFound's performance on a wide range of CXR interpretation tasks, \blue{from thoracic disease detection, opportunistic risk estimation, and malpositioned tube detection to anatomical structure segmentation}.
CheXFound outperformed previous state-of-the-art models such as RAD-DINO \cite{perez2024rad}, EVA-X \cite{yao2024eva}, and CheXzero \cite{tiu2022expert} for classifying 40 disease findings at different prevalence levels on the CXR-LT 24 dataset \cite{peng_2024_10991413}. 
Besides, CheXFound demonstrated superior label efficiency, achieving best-performing results on the Shenzhen, Montgomery, and JSRT datasets with limited training data.
% of disease  
% \rev{We evaluate ChestFound on in total}{For the long-tail thoracic disease detection tier,} 40 disease findings provided by CXR-LT Challenge\cite{holste2024towards} with different levels of disease prevalence \ins{are involved}. 
% Results show that ChestFound achieves top-3 performance on the official leaderboard with an unseen radiologist-annotated gold standard test set.
% and outperforms previous models including CheXzero\cite{tiu2022expert} and ConvNeXt\cite{liu2022convnet} Fig. \ref{fig:performance}.
We also found that CheXFound achieved significant performance increases compared with its comparisons for the out-of-distribution tasks, including opportunistic CVD risk estimation and mortality prediction on the PLCO dataset \cite{hocking2010lung}, \blue{disease detection on the VinDr-CXR dataset \cite{nguyen2022vindr}, malpositioned tube detection on the RANZCR-CLiP dataset \cite{ranzcr-clip}, and rib segmentation on the VinDr-RibCXR dataset \cite{nguyen2021vindr}}. 
Overall, we demonstrate CheXFound's strong generalization capabilities across a wide range of downstream tasks on in-distribution and out-of-distribution datasets. CheXFound's strong representation quality will enable diverse downstream adaptations with improved label efficiency in future applications.
% We further validate ChestFound's capability in cardiovascular disease risk detection and survival prediction based on a routine chest radiograph. 
% Results demonstrate that ChestFound can achieve competitive performance compared with models using computed tomography (CT) for risk prediction, highlighting the potential of ChestFound as a foundation model for a full spectrum of chest radiograph analysis tasks.

\section{Related works}
\subsection{Self-supervised Visual Representation Learning}
Our study is mostly related to self-supervised visual representation learning. After the success of masked language modeling in language domain, masked autoencoder (MAE) \cite{he2022masked} and BEiT \cite{bao2021beit} translate the idea into visual representation learning, which assume the pretext task of recovering masked pixels can train networks to learn visual information and context. Another family of self-supervised learning methods (SimCLR \cite{chen2020simple} and MoCov3\cite{chen2021empirical}) apply contrastive learning objectives, assuming augmentation invariance of image representations and aiming to learn contrastive class representations. These methods have been reported to achieve inferior linear probe performance and require fine-tuning backbone features \cite{oquab2023dinov2}. They also do not translate well into medical applications \cite{chen2024towards}. Beyond the above methods, another family of self-supervised learning methods rely on a knowledge distillation framework first introduced by BYOL \cite{grill2020bootstrap}, which bootstraps latent features of a teacher network to train a student network. DINO \cite{caron2021emerging} applies self-distillation with the Transformer architecture and enforce similarity of categorical distributions. iBOT \cite{zhou2021ibot} extends the framework with masked image modeling. DINOv2 \cite{oquab2023dinov2} carefully curates pretraining data with deduplication and further makes modifications to improve training. Overall, self-distillation methods excel at linear probe evaluation and have demonstrated generalizability in medical application \cite{perez2024rad, chen2024towards}. Our study follows this methodology to train CheXFound with strong representation quality.

% SimSiam \cite{chen2021exploring}
 % \cite{}\com{missing citation}
% \begin{itemize}
%     \item Pretext tasks: masked autoencoder (MAE), contrastive learning (SimCLR, MoCov3); limitations: fine-tuning features, weak linear probe performance, does not translate well into medical domain
%     \item Knowledege distillation: BYOL introduce teacher-student network, DINO self-distillation with [CLS] token alignment, iBOT masked image modeling, the same style as BERT
% \end{itemize}
% \subsubsection{Exponential Moving Averaged Network as Online Tokenizer}
% \subsubsection{Masked Image Modeling as Knowledge Distillation}
% \subsubsection{Self-distillation via Local-to-global Alignment}
\subsection{Foundation Models for Medical Applications}
% \blue{
% Prior to the advent of foundation models, studies used both end-to-end supervised training and supervised fine-tuning methods to develop models for chest X-ray analysis. 
% Representative works using end-to-end supervised training include CheXNet \cite{rajpurkar2017chexnet} and CheXpert \cite{irvin2019chexpert} which trained convolution neural networks for disease detection achieving better performance than radiologists. 
% In contrast, prior studies on supervised fine-tuning, such as CheXtransfer \cite{ke2021chextransfer}, MedAug \cite{vu2021medaug}, and MoCo-CXR \cite{sowrirajan2021moco}, explored different strategies: CheXtransfer directly transferred ImageNet-pretrained weights for chest X-ray analysis, while MedAug and MoCo-CXR improved the paradigm by pretraining models in the chest X-ray domain before supervised fine-tuning. 
% }
\blue{Prior to the advent of foundation models, studies used both end-to-end supervised training and supervised fine-tuning methods to develop models for chest X-ray analysis. For example, CheXNet \cite{rajpurkar2017chexnet} and CheXpert \cite{irvin2019chexpert} trained convolution neural networks from scratch, achieving better performance than radiologists for the detection of pneumonia and five selected pathologies, respectively. In contrast, supervised fine-tuning methods divided network training into two stages: pretraining to initialize network parameters and fine-tuning to adapt the network to specific tasks. Such methods include CheXtransfer \cite{ke2021chextransfer}, MedAug \cite{vu2021medaug}, and MoCo-CXR \cite{sowrirajan2021moco}. CheXtransfer directly fine-tuned a convolutional neural network pretrained on ImageNet for CXR analysis, allowing the model to effectively take advantage of natural domain knowledge to achieve enhanced performance. Both MedAug and MoCo-CXR improved the paradigm of CheXtransfer by pretraining their own models on CXR images via contrastive learning before fine-tuning them on disease detection datasets.}

The surge in available data and computational resources have enabled the large-scale pretraining of foundation models. Studies have demonstrated that scaling foundation models in data and model sizes can achieve performance increases across a wide array of downstream tasks \cite{he2022masked, oquab2023dinov2, chen2024towards}. In medical domain, studies have developed multiple categories of foundation models that use different pretraining approaches and data modalities.
% differing in technical approaches and data modalities. 
Our study belongs to the family of vision-centric foundation models. 
\blue{In the domain of CXR image analysis, several studies have explored various strategies for developing vision-centric foundation models, including RAD-DINO \cite{perez2024rad}, RayDINO \cite{moutakanni2024advancing}, EVA-X \cite{yao2024eva}, Medical MAE \cite{xiao2023delving}, Foundation Ark \cite{ma2023foundation}, and Adam \cite{hosseinzadeh2023towards}. 
However, as summarized in Table \ref{tab:vfms}, most of these related studies remain limited in either model capacity or data scale. Models such as \cite{perez2024rad, yao2024eva, xiao2023delving, ma2023foundation, hosseinzadeh2023towards} typically employed the ViT-Base or Swin Transformer-Base architecture, with up to 88M parameters, and were pretrained on up to 838K CXRs from five public datasets. To further investigate the scalability, this work scales CheXFound to the ViT-Large architecture and pretrains it on 987K CXRs from 12 different sources. Among the above CXR foundation models, only RayDINO was based on the ViT-Large architecture, as adopted in this work. However, RayDINO did not sufficiently exploit the benefit of data diversity, relying on only four commonly used public datasets for pretraining. In contrast, our work expanded the pretraining dataset with eight additional sources, resulting in 124K more pretraining images.}
% RAD-DINO \cite{perez2024rad} and EVA-X \cite{yao2024eva} are two foundation models in CXR domain. 
% Compared to CheXFound with ViT-L pretrained on CXR-1M, these models are limited in model and data scales. 
% In computational pathology, UNI \cite{chen2024towards} demonstrates generalization capabilities across downstream tasks, even for rare diseases. 

\begin{table*}[t]
\centering
\caption{\blue{Summary of the network architectures and pretraining datasets utilized by recent vision-centric foundation models}. }
\label{tab:vfms}
\makebox[\textwidth][c]{
\begin{tabular}{c c c c c}
  \toprule
  \textbf{Models} & \textbf{Architectures} & \# \textbf{Parameters}  & \# \textbf{CXRs} & \# \textbf{Data sources} \\ \midrule
  Medical MAE \cite{xiao2023delving} & ViT-Base & 86M & 509K & 3 public datasets \\
  Adam \cite{hosseinzadeh2023towards} & ResNet-50 & 25M & 224K & 1 public datasets \\
  Foundation Ark \cite{ma2023foundation} & Swin Transformer-Base & 88M & 704K & 4 public datasets \\
  EVA-X \cite{yao2024eva} & ViT-Base & 86M & 520K & 3 public datasets \\
  RAD-DINO \cite{perez2024rad} & ViT-Base & 86M & 838K & 5 public datasets + private dataset \\
  RayDINO \cite{moutakanni2024advancing} & ViT-Large & 307M & 863K & 4 public datasets \\ \midrule
  CheXFound  & ViT-Large & 307M & 987K & 12 public datasets \\
  \bottomrule
\end{tabular}
}
\end{table*}

Another category of foundation models incorporate vision and text data for multimodal pretraining. CheXzero \cite{tiu2022expert}, BiomedCLIP \cite{zhang2023biomedclip}, PubMedCLIP \cite{eslami2021does} use contrastive vision-language pretraining, which is effective in zero-shot classification. Further development of vision-language models takes advantage of instruction-tuning to improve reasoning and detailed description capabilities \cite{chen2024chexagent, li2024llava, yang2024advancing}. Overall, research empirically finds that vision-language models achieve inferior performance than vision-centric foundation models in CXR classification \cite{perez2024rad}. In this study, we focus on the vision-centric foundation model and investigate its capability for extensive CXR classification tasks.
To the best of our knowledge, our work employs the largest-scale self-supervised pretraining with over 1 million unique CXRs.
% \py{How do you distinguish your work from other works listed above?}
% surge, scaling
% , CONCH \cite{lu2024visual}, and PILP \cite{huang2023visual}

% \begin{itemize}
% \item Vision-centric foundation models; CXR: RAD-DINO, EVA-X; DINO-related vision foundation model: UNI;
% \item Vision language foundation models, contrastive vision-language pretraining: CheXzero, BiomedCLIP, PubMedCLIP, CONCH; Instruct-tuned models: CheXagent, BiomedGPT, PathChat, LLaVA-Med, Med-Gemini; versatile models: BiomedParse
% \end{itemize}

\section{Materials and Methods}

\begin{figure*}[th]
    \centering
    \includegraphics[width=0.9\textwidth]{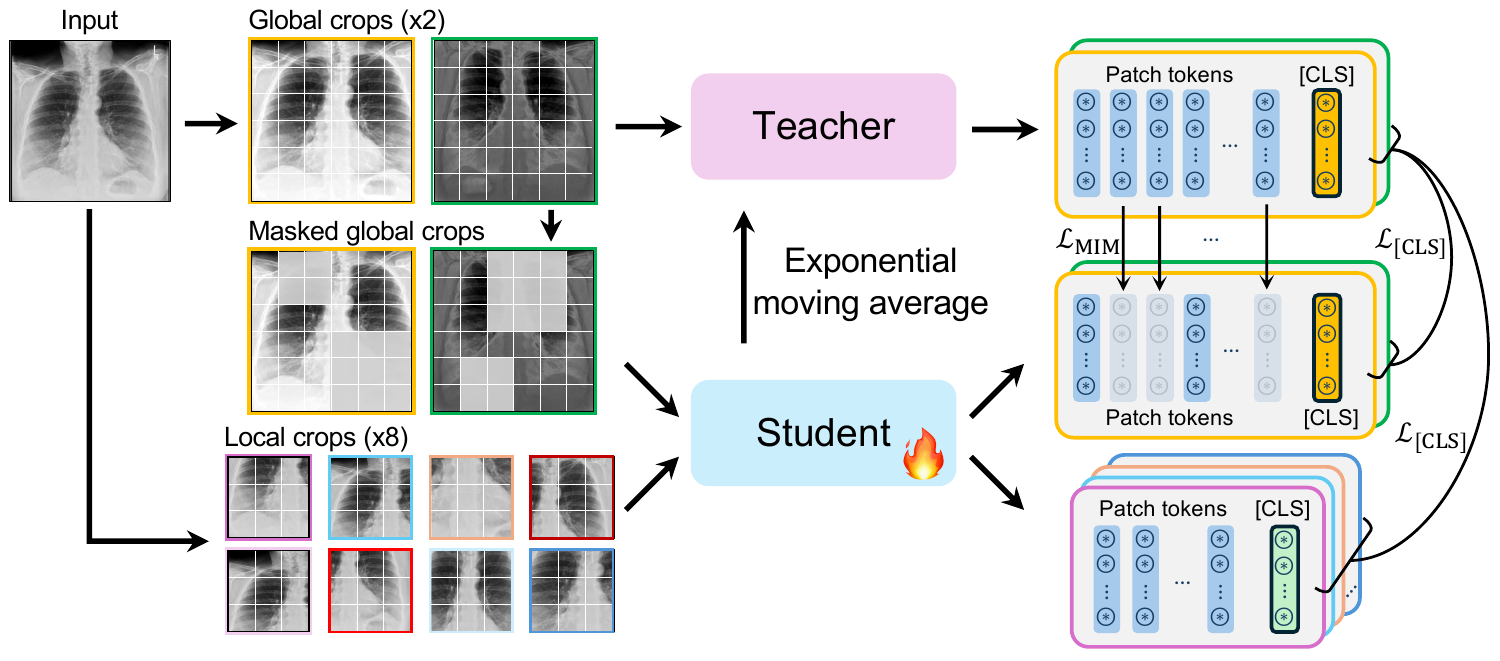}
    \caption{Overview of self-supervised pretraining of CheXFound, using publicly available CXRs from multiple institutions with a masked image modeling objective and a \texttt{[CLS]} token alignment objective.}
    \label{fig:pretrain}
\end{figure*}

\begin{table}[th]
    \setlength{\tabcolsep}{5pt}
    \setlength{\extrarowheight}{2pt}
    \centering
    \caption{Curation of CXR-987K with publicly available datasets from diverse institutions for self-supervised pretraining. CXR-987K is subsetted into CXR-207K and CXR-744K to evaluate the data scalability of self-supervised models.}
    % \com{The caption is inconsistent with the statement described in main text. It sounds like 'CXR-1M'='CXR-207K'+'CXR-744K'. BTW, I adjusted the table structure a bit. Pls review it.}
    \label{tab:ptdata}
    \begin{tabular}{l | c | c | r }
    \toprule
     \multicolumn{1}{c|}{\textbf{Datasets}} & \textbf{View} & \textbf{Findings} & \textbf{image \#} \\ \midrule
     MIMIC-CXR\cite{johnson2019mimic} & Frontal, Lateral & 14 diseases & 207,096\\
     \midrule
     \multicolumn{4}{r}{\textbf{Total number of images in CXR-207K: 207,096}}\\
     \midrule
     CheXpert\cite{irvin2019chexpert} & Frontal, Lateral & 14 diseases & 223,648\\
     PadChest\cite{bustos2020padchest} & Frontal, Lateral & 193 diseases & 160,861\\
     NIH-CXR\cite{wang2017chestx} & Frontal & 14 diseases & 112,120\\ %CXR14
     % NIH-CXR\cite{wang2017chestx} & Frontal & 14 diseases & 82,120 \\ %CXR14
     BRAX\cite{reis2022brax} & Frontal, Laterall & 14 diseases & 40,967\\ 
     \midrule
     % \multicolumn{4}{r}{\textbf{Total number of images in CXR-714K: 714,692}} \\
     \multicolumn{4}{r}{\textbf{Total number of images in CXR-744K: 744,692}}\\
     \midrule
     % VinDr-CXR\cite{nguyen2022vindr} & Frontal & 28 diseases & 18,000 \\
     CANDID-PTX\cite{feng2021curation} & Frontal & Pneumothorax & 19,237 \\
     SIIM-ACR\cite{siim-acr} & Frontal & Pneumothorax & 18,499 \\
     Object-CXR\cite{objectcxr} & Frontal & Foreign objects & 9,000\\
     COVID-19\cite{lakhani20232021} & Frontal & COVID-19 & 7,597 \\
     COVIDx CXR-4\cite{wu2023covidx} & Frontal & COVID-19 & 84,818 \\
     MIDRC COVIDx\cite{midrc-covidx} & Frontal & COVID-19 & 23,001 \\
     BIMCV COVID+\cite{vaya2020bimcv} & Frontal, Lateral & COVID-19 & 80,889 \\
     \midrule
     % \multicolumn{4}{r}{\textbf{Total number of images in CXR-1M: 1,005,733}}\\
     % \multicolumn{4}{r}{\textbf{Total number of images in CXR-957K: 957,733}}\\
     \multicolumn{4}{r}{\textbf{Total number of images in CXR-987K: 987,733}}\\
     \bottomrule
    \end{tabular}
\end{table}

%\subsection{Data Curation for CXR-1M}
\subsection{Pretraining CheXFound with Large-scale Dataset}

As detailed in Table \ref{tab:ptdata}, we curated the \blue{CXR-987K} dataset for self-supervised pretraining by retrieving in total \blue{987,733} unique CXRs from \blue{12} publicly available datasets \cite{johnson2019mimic, irvin2019chexpert, bustos2020padchest, wang2017chestx, wu2023covidx, reis2022brax, midrc-covidx, vaya2020bimcv, nguyen2022vindr, lakhani20232021, objectcxr, feng2021curation, siim-acr} that were released for various downstream tasks, including disease diagnosis, abnormality detection, foreign objection detection, and segmentation.
%by retrieving CXRs from ins{13} publicly available datasets collected for various objectives, including disease diagnosis (MIMIC-CXR \cite{johnson2019mimic}, CheXpert \cite{irvin2019chexpert}, PadChest \cite{bustos2020padchest}, CXR14 \cite{wang2017chestx}, COVIDx CXR-4 \cite{wu2023covidx}, BRAX \cite{reis2022brax}, MIDRC COVIDx \cite{midrc-covidx}, BIMCV COVID+ \cite{vaya2020bimcv}), abnormality detection (VinDr-CXR \cite{nguyen2022vindr}, COVID-19 \cite{lakhani20232021}), foreign objection detection (Object-CXR \cite{objectcxr}), and segmentation (CANDID-PTX \cite{feng2021curation}, SIIM-ACR \cite{siim-acr}), totaling 1,005,733 CXRs, as shown in Table \ref{tab:ptdata}. 
To learn comprehensive representations for multiview CXR analysis, both frontal-view in PA (posterior-anterior) or AP (anterior-posterior) and lateral-view CXRs were included into \blue{CXR-987K}. 

To evaluate the data scalability of self-supervised pretraining, we further created CXR-207K and CXR-744K, denoting two subsets of \blue{CXR-987K}, as shown in Table~\ref{tab:ptdata}. CXR-207K contains approximately 207K CXRs from MIMIC-CXR. CXR-744K contains around 744K CXRs from five datasets: MIMIC-CXR, CheXpert, PadChest, NIH-CXR, and BRAX.

We used DINOv2 \cite{oquab2023dinov2}, a state-of-the-art self-supervised learning method, to pretrain CheXFound on \blue{CXR-987K}.
DINOv2 inherits designs from DINO \cite{caron2021emerging} and iBOT \cite{zhou2021ibot} and incorporates two self-distillation objectives: the masked image modeling loss $\mathcal{L}_{\texttt{MIM}}$ and the \texttt{[CLS]} token alignment loss $\mathcal{L}_{\texttt{[CLS]}}$. It uses a teacher-student knowledge distillation architecture as shown in  Fig. \ref{fig:pretrain} to learn CXR representations. Masked image modeling uses the teacher network as an online tokenizer, which generates patch tokens from intact images to guide the student network in reconstructing masked patch tokens. This approach enables the student network to learn both visual features and contextual information effectively.
On the other hand, the \texttt{[CLS]} token alignment loss $\mathcal{L}_{\texttt{[CLS]}}$ enforces similarity between \texttt{[CLS]} tokens output by the teacher and student networks. This approach aims to train the network to learn high-level class representations with off-the-shelf linear probe capabilities.

\subsection{Global and Local Representations Integration for Multilabel Classification}
\begin{figure}
    \centering
    \includegraphics[width=\linewidth]{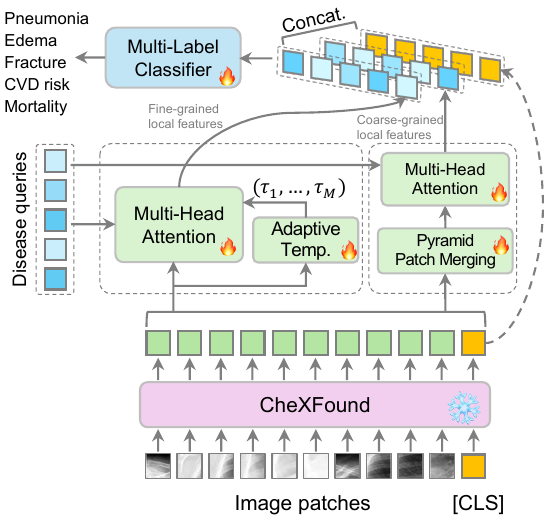}
    \caption{Global and Local Representations Integration (GLoRI) for evaluating CheXFound on downstream tasks. GLoRI is trained on top of the frozen CheXFound backbone. \blue{GLoRI incorporates adaptive temperatures and pyramid patch merging into the attention mechanism to encode fine-grained and coarse-grained local features respectively} and integrates the global image features from the [CLS] token to predict final multilabel classification results.}
    \label{fig:glori}
\end{figure}

For the downstream evaluation of CheXFound, the linear probe classifier is a pivotal tool to evaluate the quality of pretrained representations. However, the linear probe classifier has limited capability to address the multilabel classification problem commonly seen in CXR interpretation, since it generally relies on the global image features from a single \texttt{[CLS]} token for classifying a wide range of pathologies and lacks essential local details to support the predictions. In contrast, patch tokens from our pretrained CheXFound contain rich CXR representations and high-level contextual information learned via masked image modeling, which can provide disease-specific local features to substantially reduce ambiguities arisen from using the \texttt{[CLS]} token for classifying multiple pathologies. 
To take advantage of both local and global features for multilabel disease classification, we introduce GLoRI (Fig. \ref{fig:glori}), \blue{which consists of two local-feature branches that use an adaptive temperature module to provide fine-grained local features and a pyramid patch merging module to provide coarse-grained local features respectively and a skip connection to integrate the global image features from the $\texttt{[CLS]}$ token for multilabel disease classification.}
% which utilizes a cross-attention layer with disease queries to summarize patch token features and a skip-connection to integrate the $\texttt{[CLS]}$ token towards final prediction.

\subsubsection{Encoding Fine-grained Local Features with Adaptive Temperatures}
% To be specific,
GLoRI receives output patch tokens $\mathbf{u}^\texttt{Patch} \in \mathbb{R}^{N \times D_\text{model}}$ from the frozen CheXFound backbone as input, where $D_\text{model}$ is the backbone embedding dimension. Since there can be a dimension mismatch between the backbone and GLoRI, we use a linear embedding layer to project $\mathbf{u}^\texttt{Patch}$ to the GLoRI dimensional space:

\begin{equation}
\mathbf{u}^{\prime\texttt{Patch}} = \text{ReLU}(\text{Linear}^\text{embed}(\mathbf{u}^\texttt{Patch})),
\end{equation}
where $\mathbf{u}^{\prime\texttt{Patch}} \in \mathbb{R}^{N\times D_\text{GLoRI}}$ is the projected by the linear embedding layer $\text{Linear}^\text{embed}(\cdot)$ to the patch token sequence with dimension $D_\text{GLoRI}$. In GLoRI, to extract disease-specific local features using the multi-head attention layer, we initialize $M$ disease queries corresponding to $M$ disease findings, denoted as $\mathbf{q} \in \mathbb{R}^{M\times D_\text{key}}$. The keys $\mathbf{k} \in \mathbb{R}^{N\times D_\text{key}}$ and values $\mathbf{v} \in \mathbb{R}^{N\times D_\text{value}}$ for the multi-head attention layer are derived from $\mathbf{u}^{\prime\texttt{Patch}}$ to provide rich CXR representations.

\blue{Additionally, to improve GLoRI's capability to focus on small or diffuse abnormal regions, we introduce an adaptive temperature module to adjust the attention in the multi-head attention layer. The adaptive temperature module projects the average-pooled features of $\mathbf{u}^\texttt{Patch}$ through a multilayer perceptron with a tanh activation function and then applies a natural exponential function to generate the temperature vector $\boldsymbol{\uptau} \in \mathbb{R}^{M}$, where $M$ corresponds to the number of disease queries. 
We incorporate $\boldsymbol{\uptau}$ into the scaled dot-product attention mechanism to capture fine-grained local features:}
% A scaled dot-product attention module is used to compute attention-pooled features relevant to the query diseases:
\begin{equation}
\blue{\mathbf{q}^\prime = \text{Softmax} \left(\frac{\mathbf{q}\mathbf{k}^T}{\sqrt{D_\text{key}} \boldsymbol{\uptau}} \right)\mathbf{v},}
\end{equation}
% \begin{equation}
% \mathbf{q}^\prime = \text{Softmax} \left( \frac{\text{Linear}^\text{query}(\mathbf{q})\text{Linear}^\text{key}(\mathbf{k})^T}{\sqrt{D_\text{key}}} \right) \text{Linear}^\text{value}(\mathbf{v}),
% \end{equation}
% \jj{The equation of attention mechanism is not correct. (1) the linear layers are different for extracting k, q, and v. (2) the input of k and q layers should be the same. (3) Explaining the definition of Dkey may be better.} 
\blue{where $\mathbf{q}^\prime \in \mathbb{R}^{M\times D_\text{key}}$ denotes the output disease queries that capture fine-grained local features.
$\boldsymbol{\uptau}$ adjusts the distribution of attention weights for each disease query to be either sharper or smoother with a value lower or larger than 1 respectively.}
% $D_\text{key}$ is the dimension of the key tokens.

\subsubsection{Encoding Coarse-grained Local Features with Pyramid Patch Merging}
\blue{To effectively summarize contextual information, we incorporate a pyramid patch merging module to integrate multi-scale patch embeddings at low resolutions to provide coarse-grained local features. The pyramid patch merging module merges $8\times8$, $4\times4$, and $2\times2$ adjacent patch tokens in $\mathbf{u}^\texttt{Patch}$ via average pooling to generate multi-scale patch embeddings. Then, it projects these patch embeddings to a dimension of $\frac{D_\text{model}}{3}$ via a linear layer and a ReLU activation, upsamples them to the original resolution, and concatenate them to construct patch embeddings with a dimension of $D_\text{model}$. We then apply a layer normalization to the patch embeddings before inputting them to the multi-head attention to encode coarse-grained local features.}

\subsubsection{Integrating Global Image Features}
Last, we integrate the fine- and coarse-grained local features with the \texttt{[CLS]} token to construct the GLoRI output token sequence. 
For multilabel classification, each GLoRI output token represents a disease finding and is projected by a linear classifier, optimized using a binary cross-entropy loss, to perform disease detection.

\blue{Overall, GLoRI employs the attention pooling mechanism, a strategy also employed by CLIP \cite{radford2021learning}, V-JEPA \cite{bardes2024revisiting}, RayDINO \cite{moutakanni2024advancing}, and GLoRIA \cite{huang2021gloria} that incorporated attention pooling for image representation pooling \cite{radford2021learning, huang2021gloria}, video classification \cite{bardes2024revisiting}, and disease detection \cite{moutakanni2024advancing}. GLoRI distinguishes itself from these related works by integrating two local-feature branches to capture both fine- and coarse-grained local features.}
% Perceiver \cite{jaegle2021perceiver} and DETR \cite{carion2020end}, for CheXFound's downstream evaluation.
We empirically demonstrate that GLoRI can extract local features relevant to disease abnormalities in CXRs in Section \ref{sec:intepret}.

% uses disease queries to compute attention-pooled CXR representations

% For the opportunistic CXR interpretation, additional disease queries are created for GLoRI to retrieve representations related to the CVD risk estimation and mortality prediction.

% \begin{figure*}
%     \centering
%     \includegraphics[width=0.8\textwidth]{figs/framework.pdf}
%     \caption{Overview of CheXFound pretraining and GLoRI. \textbf{a}. CheXFound is pretraiend on publicly available CXRs from a range of institutions using a self-supervised training algorithm that consists of a masked image modeling objective and an alignment objective. \textbf{b}. For the evaluation of CheXFound on downstream tasks, GLoRI is appended on top of the frozen CheXFound backbone. GLoRI uses query tokens to compute attention-pooled CXR representations and integrates the global features from [CLS] token for disease finding classification. \textbf{c}. For the opportunistic CXR interpretation, additional query tokens are created for GLoRI to retrieve representations related to the estimation of CVD risk and all-cause mortality. \py{Please merge panels b and c into one single figure.}}
%     \label{fig:framework}
% \end{figure*}

% \section{Experiments and results}
\section{Experimental Design}

\subsection{Implementation details}
\subsubsection{Self-supervised pretraining details}
We conducted self-supervised pretraining of CheXFound on our curated \blue{CXR-987K} dataset (Table \ref{tab:ptdata}) using the ViT-L architecture with a patch size of 16$\times$16 pixels. 
% The categorical distribution dimension $K$ is set to 131,072 for both $\mathcal{L}_\texttt{[CLS]}$ and $\mathcal{L}_\texttt{MIM}$ objectives.
% \com{There is no 'K' defined in Equation 1 and 2. What is its definition? And the value of K looks weird. Is it a typo?} 
The loss weights for $\mathcal{L}_\texttt{[CLS]}$ and $\mathcal{L}_\texttt{MIM}$ were set to 1.0 and 3.0, respectively. The momentum to compute the exponential moving average of the student network was set to 0.994. We varied the global and local crop sizes to pretrain CheXFound at a \blue{range} of resolutions. Specifically, we set the global and local crop size pairs to be (512, 144), (448, 128), (336, 128), and (224, 96). We set the number of global and local crops to 2 and 8, respectively. For masked image modeling, we set the proportion of masked patches to the range (0.1, 0.5). We trained CheXFound for 100 epochs with an epoch length of 2,500 iterations and a batch size of 14 per graphics processing unit (GPU). We used the AdamW optimizer with an initial learning rate of 2e-4. We applied a Cosine annealing schedule for learning rate decay and a warm up period of 10 epochs. 
We pretrained CheXFound on a DGX-1 server with $8\times$ NVIDIA A100 40GB GPUs. 
% \rev{8$\times$40 GB}
Depending on the image resolutions, the pretraining processes take around 48 to 96 hours. 
% The project source code is publicly available at \url{https://github.com/RPIDIAL/CheXFound}.

\subsubsection{Downstream evaluation details}
In downstream evaluation, we trained GLoRI with feature representations from the frozen CheXFound. We took the concatenated representations from the last 4 layers of CheXFound as the input to GLoRI. We set the embedding dimension of GLoRI to 768. Disease queries were randomly initialized with a standard normal distribution. For the cross-attention layer, we used the multihead attention mechanism and set the number of heads to 8. To train GLoRI, we used the AdamW optimizer and conduct a thorough learning rate search in \{1e-5, 2e-5, 5e-5, 1e-4, 2e-4, 5e-4, 1e-3, 2e-3, 5e-3\} to obtain the best-performing learning rate on the validation set. We then combined the training and validation sets for a second round of training using the best learning rate. To maximize the number of images that the GLoRI head processes during downstream adaptation, we trained GLoRI for 10 epochs on CXR-LT 24, CheXpert, and PLCO and 100 epochs on Shenzhen, Mongomery, and JSRT. We set the batch size to 16 for Montgomery and JSRT and 256 for CXR-LT 24, CheXpert, Shenzhen, and PLCO. 

\subsection{Experimental datasets}

% \py{Did other foundation models split dataset into training/val/test and test on a portion only? If the foundation models are frozen, the computation is fairly light. Why not perform cross validation? I guess you used the training set also for pretraining. You should mention it to clarify it. How about the val set, used in pretraining? If not, it is possible to do cross validation on val+test. What is your justification?}

% in-distribution and out-of-distribution datasets
To rigorously evaluate CheXFound's in-distribution and out-of-distribution performance, we employed an extensive classification benchmark, consisting of in-distribution datasets (CXR-LT 24 \cite{peng_2024_10991413} and CheXpert \cite{irvin2019chexpert}) and out-of-distribution datasets (\blue{VinDr-CXR \cite{nguyen2022vindr}, CXR-Pneumonia \cite{kermany2018identifying}}, Shenzhen \cite{jaeger2014two}, Montgomery\cite{jaeger2014two}, Japanese Society of Radiological Technology (JSRT) \cite{shiraishi2000development}, \blue{RANZCR-CLiP \cite{ranzcr-clip}}, and Prostate, Lung, Colorectal, and Ovarian (PLCO) Cancer Screening Trial \cite{hocking2010lung}).
Following recent works in vision-centric foundation models \cite{perez2024rad, chen2024towards}, we split the evaluation datasets into training, validation, and test splits. To avoid any potential data contamination, we included only the training set in CXR-LT 24 and the training and validation sets in CheXpert for self-supervised pretraining, while keeping the test set unseen. For the out-of-distribution datasets (\blue{VinDr-CXR}, \blue{CXR-Pneumonia}, Shenzhen, Montgomery, JSRT, \blue{RANZCR-CLiP}, and PLCO), none of the images in the training, validation, and test sets were used for self-supervised pretraining.

% \com{Do we need to add these benchmarking datasets to Table I?}
Although the MIMIC-CXR dataset is the common benchmark to evaluate the performance of CXR interpretation models, its labels contain only 14 findings.
%its CXR images were released with 14 binary labels indicating the existence of pathologies. 
To assess the generalizability of foundation models across diverse disease types, we conducted experiments on the CXR-LT 24 dataset which includes the annotations of 40 disease findings at different levels of prevalence (Fig. \ref{fig:prevalence_cxrlt}a). %by extracting labels from the MIMIC-CXR radiology reports. CXR-LT 24 released labels for a total of 258,871 CXRs. 
% of 258,871 CXRs 
To evaluate the model performance, we divided CXR-LT 24 into a training set of 207,096 images and a test set of 51,775 images. 
% [Be consistent about the name, either 24 or 2024, but not both]

To evaluate the performance of the foundation models against the annotations on five selected pathologies (atelectasis, cardiomegaly, consolidation, edema, and pleural effusion) by board-certificated radiologists, we incorporated the CheXpert dataset. It was divided into 191,027 frontal-view images in the training set,  202 images in the validation set, and 518 images in the test set. 

To assess the out-of-distribution generalization capabilities of the foundation models, \blue{we evaluated model performance on VinDr-CXR that contains labels for 28 diseases. We used the official data splits of VinDr-CXR that contain 15000 CXRs for training and 3000 CXRs for test.}
We also performed evaluation on CXR-Pneumonia for pneumonia detection, Shenzhen and Montgomery for tuberculosis detection, and JSRT for lung nodule detection. 
\blue{The CXR-Pneumonia dataset contains 5,232 / 624 CXRs for training and test respectively.}
% We note that none of these images was used in the pretraining.
The Shenzhen, Montgomery, and JSRT datasets were divided into training, validation, and test splits with a ratio of 70:10:20.
Shenzhen contains 463 training images, 65 validation images, and 134 test images. Montgomery contains 96 training images, 14 validation images, and 28 test images. JSRT contains 171 training images, 24 validation images, and 50 test images.
\blue{In addition, we evaluated model performance for malpositioned tube detection on RANZCR-CLiP. We split the RANZCR-CLiP dataset into 18,742 / 2,678 / 5,356 CXRs for training, validation, and test respectively.}

To evaluate the extended predictive power of the foundation models, we obtained the lung screening CXRs from the PLCO trial and extracted the all-cause mortality and cardiovascular disease mortality labels from the up to 25-year follow-up data. The PLCO CXRs were divided into training, validation, and test sets of 133,543 images, 19,099 images, and 38,058 images respectively.

Since the CXR interpretation problem often involves severe class imbalance, we employed two metrics to evaluate model performance: the area under the precision-recall curve (AUPRC) and the area under the receiver operating characteristic curve (AUROC). 
For the multilabel classification problem, we computed the average of the metrics over disease findings.
% mean average precision (AUPRC)
% \py{Is AUPRC equivalent to AUPRC? If so, you should just use the terms of AUPRC and AUROC, which is not only easy to describe but also more commonly used in medical data analysis.}
We estimated the 95\% confidence intervals of the model performance in AUPRC and AUROC over 1,000 bootstrapped samples. To test statistical significance, we used a two-sided permutation test with 1,000 permutations to assess the observed performance differences of the two models for disease findings.

% paired
% \begin{itemize}
%     \item Mean average precision (AUPRC) metric  % useful for imbalanced datasets
%     \item Area under the receiver operating characteristic curve (AUROC) metric
%     \item Statistical significant test 
% \end{itemize}

\section{Experimental Results}

\begin{table*}
    \caption{Classification results of CheXFound and other foundation models using linear probe and GLoRI on in-distribution datasets. %presented in mean average precision (AUPRC) and area under the receiver operating characteristic curve (AUROC) over 1,000 bootstrapped samples. 
    Values inside the parentheses indicate the 95\% confidence intervals. Values in \textbf{bold} indicate the best results.}
    % \py{Pls use multirow to group `Linear probe's together. Similarly for `GLoRI head'.}
    % \py{Where do you introduce bootstrapping? I couldn't find it in the paper.}
    \label{tab:overall}
    \setlength{\tabcolsep}{2pt}
    \setlength{\extrarowheight}{2pt}
    \centering
    \begin{scriptsize}
    \begin{tabular}{c | c | c c | c c}
    \toprule
    \textbf{Classifier} & \textbf{Foundation} & \multicolumn{2}{c}{\textbf{CXR-LT 24}} & \multicolumn{2}{c}{\textbf{CheXpert}} \\ \cmidrule(lr){3-6}
    \textbf{Methods} & \textbf{Models} & \textbf{AUPRC} & \textbf{AUROC} & \textbf{AUPRC} & \textbf{AUROC} \\ \midrule
     \multirow{6}{*}{Linear probe} & PubMedCLIP \cite{eslami2021does} & 0.089\tiny(.088--.089) & 0.561\tiny(.554--.568) & 0.277\tiny(.244--.311) & 0.595\tiny(.564--.628) \\
     & BiomedCLIP \cite{zhang2023biomedclip} & 0.117\tiny(.116--.118) & 0.643\tiny(.636--.649) & 0.557\tiny(.504--.607) & 0.841\tiny(.822--.860) \\
     & CheXzero \cite{tiu2022expert} & 0.112\tiny(.111--.112) & 0.552\tiny(.545--.558) & 0.468\tiny(.424--.509) & 0.778\tiny(.754--.798) \\
     & EVA-X \cite{yao2024eva} & 0.114\tiny(.113--.115) & 0.596\tiny(.590--.602) & 0.468\tiny(.422--.512) & 0.788\tiny(.763--.812) \\
     & RAD-DINO \cite{perez2024rad} & 0.114\tiny(.113--.114) & 0.557\tiny(.570--.583) & 0.463\tiny(.422--.503) & 0.746\tiny(.715--.778) \\
     & CheXFound & 0.209\tiny(.204--.214) & 0.799\tiny(.794--.804) & 0.620\tiny(.630--.727) & 0.876\tiny(.860--.892) \\ \midrule
     \multirow{6}{*}{GLoRI head} & PubMedCLIP \cite{eslami2021does} & 0.116\tiny(.115--.117) & 0.649\tiny(.643--.655) & 0.501\tiny(.450--.552) & 0.804\tiny(.778--.828) \\
     & BiomedCLIP \cite{zhang2023biomedclip} & 0.122\tiny(.121--.123) & 0.643\tiny(.636--.649) & 0.552\tiny(.506--.593) & 0.829\tiny(.809--.847) \\
     & CheXzero \cite{tiu2022expert} & 0.131\tiny(.130--.132) & 0.671\tiny(.665--.677) & 0.599\tiny(.551--.647) & 0.888\tiny(.868--.905)  \\
     & EVA-X \cite{yao2024eva} & 0.149\tiny(.147--.150) & 0.679\tiny(.672--.685) & 0.614\tiny(.571--.659) & 0.870\tiny(.853--.888) \\
     & RAD-DINO \cite{perez2024rad} & 0.173\tiny(.171--.176) & 0.723\tiny(.717--.729) & 0.639\tiny(.593--.687) & 0.884\tiny(.869--.898) \\
     & CheXFound & \textbf{0.265}\tiny(.259--.271) & \textbf{0.840}\tiny(.836--.844) & \textbf{0.679}\tiny(.630--.727) & \textbf{0.908}\tiny(.894--.921)  \\
    \bottomrule
    \end{tabular}
    \end{scriptsize}
\end{table*}

% \textbf{0.252}\tiny(.247--.258)
% \textbf{0.830}\tiny(.826--.834)

\begin{table*}
    \caption{\blue{Classification results of CheXFound and other foundation models on out-of-distribution datasets. %presented in mean average precision (AUPRC) and area under the receiver operating characteristic curve (AUROC) over 1,000 bootstrapped samples. 
    Values inside the parentheses indicate the 95\% confidence intervals. Values in \textbf{bold} indicate the best results.}}
    % \py{Pls use multirow to group `Linear probe's together. Similarly for `GLoRI head'.}
    % \py{Where do you introduce bootstrapping? I couldn't find it in the paper.}
    \label{tab:ood}
    \setlength{\tabcolsep}{1.5pt}
    \setlength{\extrarowheight}{2pt}
    \centering
    \begin{scriptsize}
    \begin{tabular}{c | c | c c | c c | c c | c c | c c}
    \toprule
    \textbf{Classifier} & \textbf{Foundation} & \multicolumn{2}{c}{\textbf{\blue{VinDr-CXR}}} & \multicolumn{2}{c}{\textbf{\blue{CXR-Pneumonia}}} & \multicolumn{2}{c}{\textbf{Shenzhen}} & \multicolumn{2}{c}{\textbf{Montgomery}} & \multicolumn{2}{c}{\textbf{JSRT}} \\ \cmidrule(lr){3-12}
    \textbf{Methods} & \textbf{Models} & \textbf{AUPRC} & \textbf{AUROC} & \textbf{AUPRC} & \textbf{AUROC} & \textbf{AUPRC} & \textbf{AUROC} & \textbf{AUPRC} & \textbf{AUROC} & \textbf{AUPRC} & \textbf{AUROC} \\ \midrule
     \multirow{6}{*}{Linear probe} & PubMedCLIP \cite{eslami2021does} & 0.088\tiny(.082--.095) & 0.570\tiny(.514--.609) & 0.954\tiny(.936--.969) & 0.924\tiny(.904--.944) & 0.857\tiny(.786--.913) & 0.814\tiny(.738--.887) & 0.565\tiny(.306--.817) & 0.534\tiny(.310--.750) & 0.785\tiny(.630--.912) & 0.685\tiny(.511--.833)\\
     & BiomedCLIP \cite{zhang2023biomedclip} & 0.100\tiny(.094--.106) & 0.587\tiny(.544--.618) & 0.984\tiny(.976--.990) & 0.973\tiny(.962--.986) & 0.903\tiny(.843--.949) & 0.885\tiny(.827--.934) & 0.925\tiny(.791--1.0) & 0.929\tiny(.822--1.0) & 0.589\tiny(.427--.755) & 0.432\tiny(.265--.601)\\
     & CheXzero \cite{tiu2022expert} & 0.102\tiny(.096--.109) & 0.536\tiny(.486--.567) & 0.973\tiny(.962--.981) & 0.953\tiny(.938--.967) & 0.929\tiny(.884--.962) & 0.906\tiny(.851--.950) & 0.964\tiny(.869--1.0) & 0.970\tiny(.898--1.0) & 0.759\tiny(.597--.885) & 0.620\tiny(.453--.777)\\
     & EVA-X \cite{yao2024eva} & 0.105\tiny(.099--.111) & 0.542\tiny(.503--.572) & 0.975\tiny(.959--.988) & 0.968\tiny(.953--.980) & 0.840\tiny(.738--.921) & 0.824\tiny(.746--.898) & 0.577\tiny(.320--.809) & 0.509\tiny(.278--.749) & 0.641\tiny(.472--.797) & 0.490\tiny(.312--.660)\\
     & RAD-DINO \cite{perez2024rad} & 0.102\tiny(.096--.107) & 0.477\tiny(.441--.507) & 0.995\tiny(.991--.998) & 0.992\tiny(.986--.996) & 0.883\tiny(.818--.933) & 0.861\tiny(.798--.916) & 0.637\tiny(.369--.842) & 0.561\tiny(.316--.788) & 0.747\tiny(.585--.888) & 0.623\tiny(.456--.778)\\
     & CheXFound & 0.298\tiny(.280--.317) & 0.838\tiny(.776--.866) & \textbf{0.996\tiny(.993--.998)} & \textbf{0.993\tiny(.988--.997)} & 0.974\tiny(.949--.992) & 0.967\tiny(.935--.990) & 0.988\tiny(.939--1.0) & 0.990\tiny(.952--1.0) & 0.918\tiny(.826--.975) & 0.856\tiny(.741--.948)\\ \midrule
     \multirow{6}{*}{GLoRI head} & PubMedCLIP \cite{eslami2021does} & 0.104\tiny(.098--.113) & 0.589\tiny(.532--.640) & 0.973\tiny(.961--.983) & 0.960\tiny(.944--.973) & 0.897\tiny(.832--.946) & 0.867\tiny(.807--.927) & 0.628\tiny(.363--.866) & 0.694\tiny(.484--.872) & 0.652\tiny(.483--.810) & 0.493\tiny(.330--.654)\\
     & BiomedCLIP \cite{zhang2023biomedclip} & 0.094\tiny(.089--.102) & 0.537\tiny(.497--.566) & 0.987\tiny(.981--.992) & 0.978\tiny(.969--.986) & 0.921\tiny(.870--.957) & 0.897\tiny(.840--.944) & 0.900\tiny(.738--1.0) & 0.898\tiny(.744--1.0) & 0.634\tiny(.470--.789) & 0.505\tiny(.291--.704)\\
     & CheXzero \cite{tiu2022expert} & 0.074\tiny(.072--.078) & 0.489\tiny(.446--.515) & 0.982\tiny(.972--.989) & 0.972\tiny(.959--.982) & 0.912\tiny(.852--.954) & 0.894\tiny(.838--.942) & 0.726\tiny(.451--.901) & 0.653\tiny(.396--.882) & 0.640\tiny(.467--.807) & 0.485\tiny(.323--.662)\\
     & EVA-X \cite{yao2024eva} & 0.163\tiny(.155--.172) & 0.662\tiny(.610--.694) & 0.990\tiny(.979--.997) & 0.988\tiny(.978--.995) & 0.928\tiny(.881--.964) & 0.896\tiny(.835--.945) & 0.977\tiny(.909--1.0) & 0.979\tiny(.918--1.0) & 0.866\tiny(.760--.936) & 0.748\tiny(.615--.864) \\
     & RAD-DINO \cite{perez2024rad} & 0.155\tiny(.147--.163) & 0.680\tiny(.630--.718) & 0.994\tiny(.991--.997) & 0.991\tiny(.985--.995) & 0.909\tiny(.854--.952) & 0.885\tiny(.823--.936) & 0.909\tiny(.747--1.0) & 0.911\tiny(.782--1.0) & 0.683\tiny(.503--.861) & 0.615\tiny(.440--.783)\\
     & CheXFound & \textbf{0.341}\tiny(.319--.367) & \textbf{0.872}\tiny(.813--.898) & \textbf{0.996\tiny(.992--.998)} & \textbf{0.993\tiny(.987--.997)} & \textbf{0.983}\tiny(.960--.996) & \textbf{0.978}\tiny(.951--.995) & \textbf{1.000}\tiny(1.0--1.0) & \textbf{1.000}\tiny(1.0--1.0) & \textbf{0.986}\tiny(.956--1.0) & \textbf{0.975}\tiny(.931--1.0)\\
    \bottomrule
    \end{tabular}
    \end{scriptsize}
\end{table*}

\begin{table*}
    \caption{Comparison of CheXFound with the end-to-end trained model and vision-language foundation models.
    %presented in mean average precision (AUPRC) and area under the receiver operating characteristic curve (AUROC) over 1,000 bootstrapped samples. 
    Values inside the parentheses indicate the 95\% confidence intervals. Values in \textbf{bold} indicate the best results.}
    % \py{Pls use multirow to group `Img-text align's together.}
    \label{tab:compare_vlm}
    \setlength{\tabcolsep}{1.2pt}
    \setlength{\extrarowheight}{2pt}
    \centering
    \begin{scriptsize}
    \begin{tabular}{c | c | c c | c c | c c | c c | c c}
    \toprule
    \textbf{Classifier} & \textbf{Foundation} & \multicolumn{2}{c}{\textbf{CXR-LT 24}} & \multicolumn{2}{c}{\textbf{CheXpert}} & \multicolumn{2}{c}{\textbf{Shenzhen}} & \multicolumn{2}{c}{\textbf{Montgomery}} & \multicolumn{2}{c}{\textbf{JSRT}} \\ \cmidrule(lr){3-12}
     \textbf{Methods} & \textbf{Models} & \textbf{AUPRC} & \textbf{AUROC} & \textbf{AUPRC} & \textbf{AUROC} & \textbf{AUPRC} & \textbf{AUROC} & \textbf{AUPRC} & \textbf{AUROC} & \textbf{AUPRC} & \textbf{AUROC} \\ \midrule
      End-to-end & ConvNeXt \cite{liu2022convnet} & 0.170\tiny(.168--.173) & 0.761\tiny(.756--.766) & 0.657\tiny(.608--.704) & 0.886\tiny(.867--.903) & 0.923\tiny(.870--.963) & 0.891\tiny(.823--.945) & 0.622\tiny(.362--.862) & 0.670\tiny(.458--.872) & 0.750\tiny(.586--.886) & 0.608\tiny(.456--.761) \\ \midrule
      \multirow{3}{*}{Img-text align.} & PubMedCLIP \cite{eslami2021does} & 0.068\tiny(.068--.069) & 0.531\tiny(.524--.537) & 0.255\tiny(.221--.294) & 0.582\tiny(.550--.618) & 0.577\tiny(.457--.700) & 0.540\tiny(.437--.637) & 0.440\tiny(.246--.672) & 0.426\tiny(.216--.658) & 0.596\tiny(.426--.758) & 0.403\tiny(.244--.572) \\
      & BiomedCLIP \cite{zhang2023biomedclip} & 0.071\tiny(.071--.072) & 0.539\tiny(.533--.545) & 0.356\tiny(.317--.393) & 0.653\tiny(.626--.681) & 0.795\tiny(.705--.878) & 0.760\tiny(.680--.836) & 0.786\tiny(.567--.936) & 0.714\tiny(.487--.912) & 0.692\tiny(.516--.859) & 0.554\tiny(.397--.722) \\
      & CheXzero \cite{tiu2022expert} & 0.134\tiny(.133--.136) & 0.668\tiny(.662--.674) & 0.646\tiny(.600--.692) & 0.888\tiny(.868--.905) & 0.875\tiny(.804--.933) & 0.849\tiny(.776--.911) & 0.967\tiny(.870--1.0) & 0.969\tiny(.889--1.0) & 0.708\tiny(.532--.863) & 0.530\tiny(.378--.698) \\ \midrule
      GLoRI head & CheXFound & \textbf{0.265}\tiny(.259--.271) & \textbf{0.840}\tiny(.836--.844) & \textbf{0.679}\tiny(.630--.727) & \textbf{0.908}\tiny(.894--.921) & \textbf{0.983}\tiny(.960--.996) & \textbf{0.978}\tiny(.951--.995) & \textbf{1.000}\tiny(1.0--1.0) & \textbf{1.000}\tiny(1.0--1.0) & \textbf{0.986}\tiny(.956--1.0) & \textbf{0.975}\tiny(.931--1.0) \\
    \bottomrule
    \end{tabular}
    \end{scriptsize}
\end{table*}

\subsection{Overall performance comparison}
% independent heads for each datasets
% ablation study that combine the datasets (CXR-LT, CheXpert) for training
% \begin{itemize}
%     \item Perform plug-and-play bootstrapping from multiple frozen image encoder
% \end{itemize}
A pivotal characteristic of foundation models lies in their capability to achieve improved performance on a wide range of downstream datasets. To evaluate the capability of foundation models, we compared CheXFound, which uses ViT-L pretrained on \blue{CXR-987K}, with publicly available pretrained encoders, including RAD-DINO \cite{perez2024rad}, EVA-X \cite{yao2024eva}, CheXzero \cite{tiu2022expert}, BiomedCLIP \cite{zhang2023biomedclip}, PubmedCLIP \cite{eslami2021does}, and ConvNeXt \cite{liu2022convnet}. 
RAD-DINO was pretrained on a combined dataset comprising MIMIC-CXR, CheXpert, PadChest, NIH-CXR, and BRAX, using the DINOv2 framework. EVA-X was pretrained on MIMIC-CXR, CheXpert, and NIH-CXR using EVA \cite{fang2023eva} technique with contrastive vision features for masked image modeling. CheXzero, a vision-language foundation model, was pretrained on MIMIC-CXR using contrastive language-image pretraining (CLIP) \cite{radford2021learning}. BiomedCLIP was pretrained on PMC-15M with image-text pairs collected from scientific articles. PubmedCLIP was initialized with the CLIP model and finetuned on PubMed articles. Finally, ConvNeXt was \blue{pretrained} on the ImageNet-22K dataset. 

\begin{figure}
    \centering
    \includegraphics[width=\linewidth]{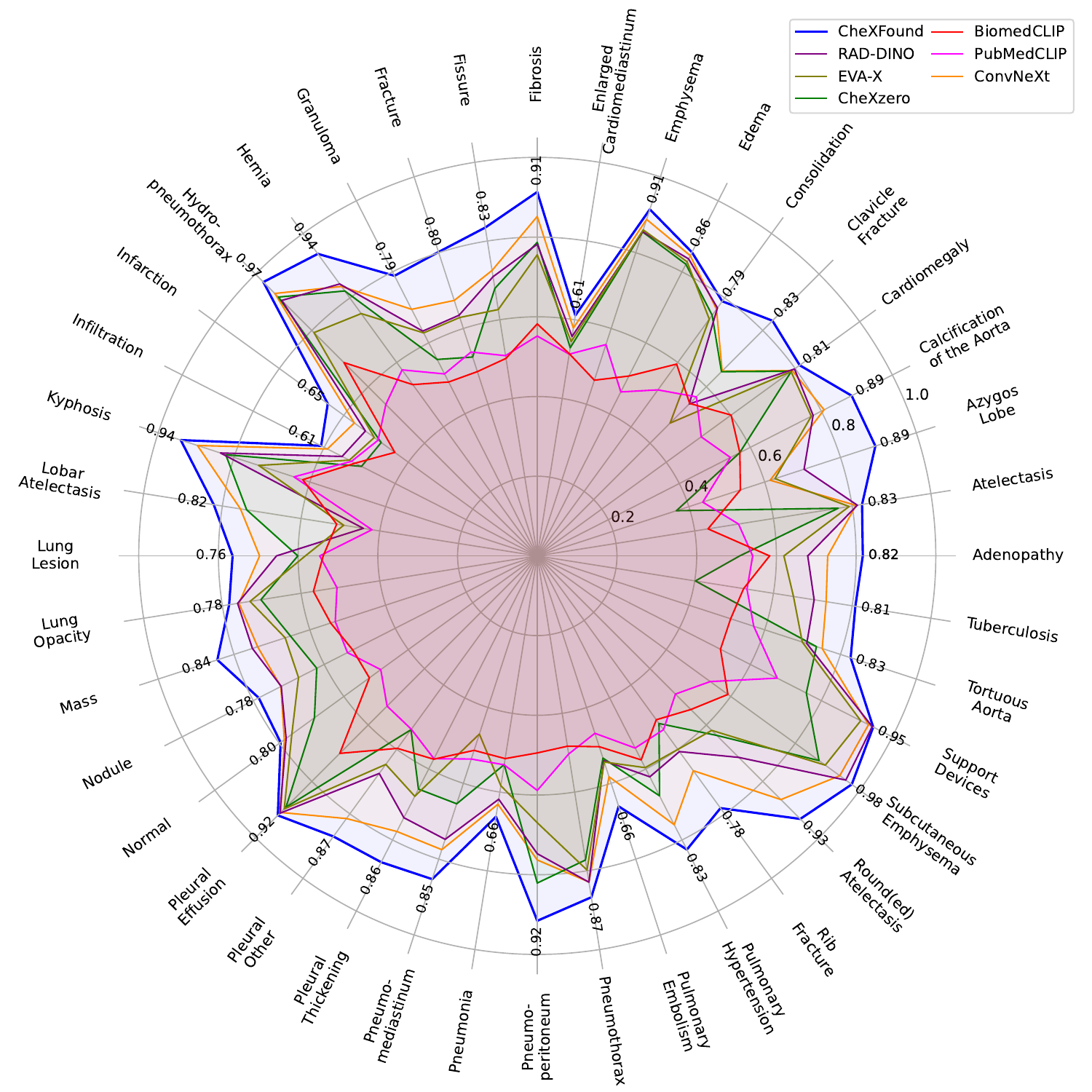}
    \caption{Detailed Performance for 40 disease findings in AUROC on the CXR-LT 24 dataset. Our CheXFound is compared with the vision-centric foundation models (EVA-X and RAD-DINO), the vision-language pretrained foundation models (CheXzero, BiomedCLIP, and PubMedCLIP), and the end-to-end trained model (ConvNeXt) with ImageNet-22K pretraining.}
    \label{fig:radar_cxrlt}
\end{figure}

\begin{figure}[t]
    \centering
    \includegraphics[width=\linewidth]{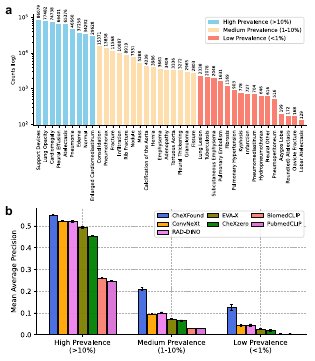}
    \caption{Model performance under high, medium and low disease prevalence. \textbf{a}, The number of labels for the 40 disease findings on the CXR-LT 24 dataset \cite{peng_2024_10991413}. \textbf{b}, Model performance in AUPRC stratified by high, medium and low disease prevalence. Error bars indicate the 95\% confidence intervals of AUPRC over 1,000 bootstrapped samples.}
    \label{fig:prevalence_cxrlt}
\end{figure}

% We investigated the capabilities of CheXFound on five publicly available CXR datasets, provided by different institutions. These include CXR-LT 24 \cite{peng_2024_10991413}, which captures 40 disease findings with a long-tail distribution as shown in Fig. \ref{fig:prevalence_cxrlt}a; CheXpert \cite{irvin2019chexpert}, containing five radiologist-annotated pathologies (atelectasis, cardiomegaly, consolidation, edema, and pleural effusion); Shenzhen and Montgomery \cite{jaeger2014two} for tuberculosis detection; and JSRT \cite{shiraishi2000development} designed for lung nodule detection. \py{This paragraph is repeating the content under experimental design. Please merge into there.}

To evaluate the effectiveness and generalizability of the representations extracted by the foundation models, we evaluated the linear probe performance across \blue{seven} datasets (CXR-LT 24, CheXpert, \blue{VinDr-CXR, CXR-Pneumonia,} Shenzhen, Montgomery, and JSRT). While linear probe stands for a straightforward approach to evaluate the quality of representations, it primarily relies on global image features from the \texttt{[CLS]} token, often resulting in suboptimal performance. We hence evaluated the pretrained encoders using our proposed GLoRI, which incorporates attention-pooled local features in addition to the global image features to enhance multilabel classification. Regarding vision-language \blue{pretrained} encoders, we further validated their vision representation quality by assessing their correlation with text features to perform zero-shot classification.

% \py{Table 2 is hard to interpret, its organization doesn't match the description below. For example, if you want to talk about linear probe first, then your Table 2 should show the performance under linear probing together. Otherwise, readers have to scan the table multiple times to collect the needed information when reading this paragraph.}
Across all the in-distribution and out-of-distribution classification tasks (CXR-LT 24, CheXpert, \blue{VinDr-CXR, CXR-Pneumonia,} Shenzhen, Montgomery, and JSRT), CheXFound consistently outperformed other foundation models in the setting using a simple linear probe classifier, as shown in Tables \ref{tab:overall} and \ref{tab:ood}. 
% \py{Is frozen backbone equal to pretrained encoder as in the above paragraph? Keep the terms consistent.} 
On the multilabel, long-tailed classification task (CXT-LT 24), CheXFound achieved an AUPRC of 0.209, outperforming the next best-performing model (either RAD-DINO or EVA-X) by 9.5\% ($p<0.001$, two-sided permutation test).
On the five-class multilabel classification task (CheXpert), CheXFound outperformed the next best-performing model (BiomedCLIP) by 3.5\% ($p<0.001$) in AUROC. \blue{On the out-of-distribution multilabel classification task (VinDr-CXR), CheXFound outperformed the next best-performing model (BiomedCLIP) by 25.1\% ($p<0.001$) in AUROC}. On out-of-distribution single-class classification tasks (\blue{CXR-Pneumonia}, Shenzhen, Montgomery, and JSRT) with limited amounts of training data, CheXFound outperformed the next best-performing models in AUROC by 0.1\% ($p>0.05$), 6.1\% ($p<0.001$), 2.0\% ($p>0.05$), and 17.1\% ($p<0.001$), respectively.
% (EVA-X) by 8.8\%  ($p<0.001$)
% \py{What are these tasks? Specify them to avoid confusion.}
% \py{Is 0.209 good or bad? Give readers a context, for example the range of AUPRC, in what range a method would be considered good.}
% \py{what are these tasks?} across the three institutions

% \py{Alternatively, change how you describe the results to match Table 2. Ask yourself the question: What is the most important thing to illustrate through Table 2? That should be at the highest level. Then the second priority, etc.}
We further evaluated the performance of the foundation models using GLoRI across the \blue{seven} classification tasks (Tables \ref{tab:overall} and \ref{tab:ood}). The foundation models (CheXFound, RAD-DINO, EVA-X, CheXzero, BiomedCLIP, and PubMedCLIP) with GLoRI generally outperformed their linear probe baselines. CheXFound with GLoRI outperformed its linear probe baseline in AUROC by 4.1\%, 3.2\%, 3.4\%, 1.1\%, 1.0\%, and 11.9\% on CXR-LT 24, CheXpert, VinDr-CXR, Shenzhen, Montgomery, and JSRT, respectively. 
CheXFound achieved an AUROC of 0.993 on CXR-Pneumonia, on par with its linear probe baseline.
In addition, as shown in Table \ref{tab:compare_vlm}, CheXFound with GLoRI outperformed CLIP-based foundation models (CheXzero, BioMedCLIP, and PubMedCLIP) when image-text alignment was used for zero-shot classification. CheXFound also outperformed the end-to-end trained model ConvNeXt. 

% \py{Justify to readers why do you need the comparison below. Use your figure well to convey more information, but not just nice to have.}
To show the detailed performance of CheXFound over 40 disease findings, we illustrated CheXFound performance in AUROC against its comparisons in Fig. \ref{fig:radar_cxrlt}. CheXFound consistently outperformed other methods across the 40 disease findings in AUROC. We also compared performance in AUPRC over disease findings with high, medium, and low prevalence in Fig. \ref{fig:prevalence_cxrlt}. CheXFound outperformed its comparisons under all levels of prevalence, even for underrepresented pathologies in the low prevalence category.

\subsection{Malpositioned tube detection}
\blue{The RANZCX-CLiP dataset contains CXRs for verifying the position of lines and tubes. This dataset categorized these CXRs into normal, borderline, and abnormal classes based on the position of tubes (endotracheal tube, nasogastric tube, and central venous catheter) to indicate whether they were appropriately positioned. The percentages of normal, borderline, and abnormal samples are 59.4\%, 28.8\%, and 11.8\% respectively, representing an unbalanced class distribution. As shown in Table \ref{tab:rancxr-clip}, CheXFound achieved AUPRCs of 0.857, 0.529, and 0.445 for classifying normal, borderline, and abnormal tubes respectively. CheXFound outperformed the next best-performing models by 16.4\% ($p < 0.001$), 16.1\% ($p < 0.001$), and 29.2\% ($p < 0.001$) for classifying normal, borderline, and abnormal classes respectively. These performance gains could be attributed to CheXFound's capability to encode high-resolution CXRs and produce effective representations of nuanced structures.} 

\begin{table}[t]
    \centering
    \setlength{\tabcolsep}{4pt}
    \setlength{\extrarowheight}{2pt}
    \caption{\blue{Classification results of foundation models for malpositioned tube detection in CXRs. We computed the mean AUPRC values and their 95\% confidence intervals over 1,000 bootstrapped samples. Values in \textbf{bold} indicate the best-performing results.}}
    \label{tab:rancxr-clip}
    \begin{tabular}{c | c c c}
        \toprule
        \textbf{Methods} & \textbf{Normal} & \textbf{Borderline} & \textbf{Abnormal} \\ \midrule
        PubMedCLIP \cite{eslami2021does}  & 0.661\tiny(.642--.679) & 0.348\tiny(.328--.372) & 0.136\tiny(.123--.149) \\
        BiomedCLIP \cite{zhang2023biomedclip}  & 0.645\tiny(.626--.664) & 0.330\tiny(.311--.350) & 0.133\tiny(.121--.148) \\
        CheXzero \cite{tiu2022expert} & 0.693\tiny(.674--.711) & 0.364\tiny(.341--.386) & 0.153\tiny(.139--.173) \\
        EVA-X \cite{oquab2023dinov2} & 0.657\tiny(.637--.676) & 0.339\tiny(.317--.361) & 0.135\tiny(.122--.150) \\
        RAD-DINO \cite{oquab2023dinov2} & 0.690\tiny(.670--.709) & 0.368\tiny(.347--.391) & 0.138\tiny(.125--.154) \\ \midrule
        CheXFound & \textbf{0.857\tiny(.844--.869)} & \textbf{0.529\tiny(.503--.556)} & \textbf{0.445\tiny(.403--.488)} \\
        \bottomrule
    \end{tabular}
\end{table}

\subsection{Opportunistic predictive power}

Beyond thoracic disease detection tasks, we evaluated \blue{CheXFound's} generalizability in opportunistic CXR interpretation. For this purpose, we requested access to the CXR arm of the PLCO trial \cite{oken2011screening}, which includes digitally scanned CXR films and up to 25-year \blue{mortality} follow-up data. Using this dataset, we investigated CheXFound's predictive capability for cardiovascular disease (CVD) risk and all-cause mortality estimation. 
%For CVD risk estimation, we predict the probability of CVD mortality. The underlying causes of death for CVD mortality includes the ischemic heart disease, cerebrovascular accident, and other circulatory disease. 
% The causes of deaths for all-cause mortality further include the respiratory illness, digestive disease, infectious disease, lung, among others. 
We used CheXFound with ViT-L pretrained on \blue{CXR-987K} in this experiment and compared CheXFound against two vision foundation models (RAD-DINO, EVA-X) and the end-to-end trained model, ConvNeXt.

\begin{figure}[t]
    \centering
    \includegraphics[width=0.8\linewidth]{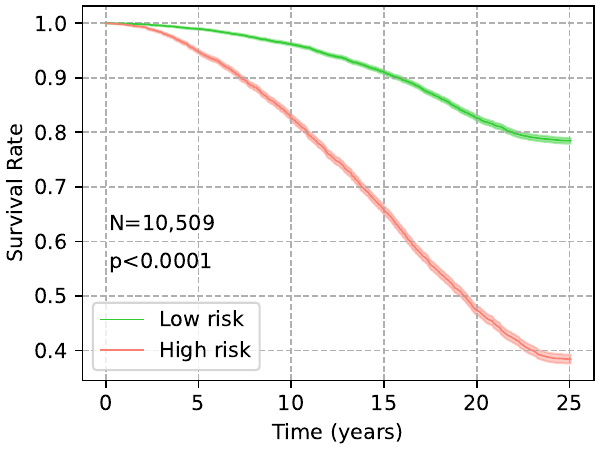}
    \caption{Kaplan-Meier curves for low-risk and high-risk groups of all-cause mortality on the PLCO dataset. The number of subjects in the test set is 10,509. The survival rates for low-risk and high-risk groups are significantly different (p$<$0.0001, log-rank test). Shaded areas indicate 95\% confidence intervals.}
    \label{fig:kaplan-meier}
\end{figure}

CheXFound consistently outperformed its counterparts in both CVD risk and all-cause mortality estimation tasks. CheXFound achieved 0.749 for CVD risk estimation and 0.786 for all-cause mortality estimation in AUROC, significantly outperforming the next best-performing method (ConvNeXt) by 3.5\% ($p<0.001$) and 4.1\% ($p<0.001$), respectively. For the all-cause mortality estimation task, we divided the test cohort into low-risk and high-risk groups based on the model prediction and computed their Kaplan-Meier curves (Fig. \ref{fig:kaplan-meier}). The survival distributions for the low-risk and high-risk groups are statistically different ($p<0.001$, log-rank test). The end-point survival probabilities are also different by a large margin for low-risk and high-risk groups (78.4\% versus 38.4\%). Overall, we demonstrated CheXFound generalization capability for opportunistic CXR interpretation.

\begin{table}[t]
    \centering
    \setlength{\tabcolsep}{1pt}
    \setlength{\extrarowheight}{2pt}
    \caption{Model performance on CVD risk and all-cause mortality estimation. Values inside the parentheses are 95\% confidence intervals. Values in \textbf{bold} indicate the best-performing results.}
    \label{tab:plco}
    \begin{tabular}{c | c c | c c}
        \toprule
        \multirow{2}{*}{\textbf{Methods}} & \multicolumn{2}{c}{\textbf{CVD risk}} & \multicolumn{2}{c}{\textbf{All-cause mortality}} \\ \cmidrule(lr){2-5}
        & \textbf{AUPRC} & \textbf{AUROC} & \textbf{AUPRC} & \textbf{AUROC} \\ \midrule
        ConvNeXt \cite{liu2022convnet} & 0.249\tiny(.240--.257) & 0.714\tiny(.705--.723) & 0.638\tiny(.629--.646) & 0.745\tiny(.736--.753) \\
        EVA-X \cite{yao2024eva} & 0.179\tiny(.171--.188) & 0.643\tiny(.635--.652) & 0.545\tiny(.536--.554) & 0.680\tiny(.674--.686) \\
        RAD-DINO \cite{perez2024rad} & 0.223\tiny(.215--231) & 0.687\tiny(.679--.695) & 0.615\tiny(.607--.622) & 0.723\tiny(.716--.729) \\ \midrule
        CheXFound & \textbf{0.289}\tiny(.276--.301) & \textbf{0.749}\tiny(.741--.756) & \textbf{0.695}\tiny(.687--.702) & \textbf{0.786}\tiny(.782--.791) \\
        \bottomrule
    \end{tabular}
\end{table}

% \py{What is the difference between AUPRC and AUPRC in the table?} \zf{CXR-LT 24 and CheXpert have multiple labels. The prefix m means macro average.}
% \begin{itemize}
%     \item CVD mortality on PLCO dataset
%     \item All-cause mortality on PLCO dataset
% \end{itemize}

% \subsection{Comparison with frozen pretrained features from open-source foundation models}
% \begin{itemize}
%     \item EVA-X
%     \item RAD-DINO
%     \item CheXagent
% \end{itemize}

% \subsection{Label efficiency of few-shot classification}
% We additionally evaluated CheXFound few-shot learning performance. Few-shot learning is an evaluation scheme to investigate the generalization capabilities of the pretrained encoders given a limited number of samples per class. For all pretrained encoders, we trained a linear probe classifier with the number of sample(s) per class $K\in\{1, 2, 4, 8, 16, 32\}$.

\subsection{CXR foundation models for semantic segmentation}
\blue{In addition to the classification tasks, we applied the pretrained CheXFound for semantic segmentation on CXRs. We included the VinDr-RibCXR \cite{nguyen2021vindr} dataset which contains segmentation masks for 10 left ribs and 10 right ribs and the Montgomery dataset \cite{jaeger2014two} which contains segmentation masks for left and right lungs. We trained a UPerNet decoder \cite{xiao2018unified} on top of the frozen foundation models to perform segmentation.
We compared CheXFound with other vision foundation models (EVA-X \cite{yao2024eva} and RAD-DINO \cite{perez2024rad}) and an end-to-end trained model (ConvNeXt \cite{liu2022convnet}). As shown in Table \ref{tab:seg}, CheXFound achieved a Dice similarity coefficient of 0.793 on VinDr-RibCXR, outperforming the next best-performing model (ConvNeXt) by 7.1\% ($p<0.001$). On Montgomery, CheXFound achieved a Dice similarity coefficient of 0.980, slightly surpassing the next best-performing model (RAD-DINO) by 0.2\%. Such marginal performance gain over RAD-DINO on Montgomery likely reflects the relative simplicity of lung segmentation on this dataset, where all comparison methods achieved satisfactory performance. In contrast, the significant performance improvements in individual rib segmentation on VinDr-RibCXR demonstrates CheXFound’s superior capability to adapt to complicated anatomical structure segmentation tasks.
}

\begin{table}[t]
    \centering
    \setlength{\tabcolsep}{1.5pt}
    \setlength{\extrarowheight}{2pt}
    \caption{\blue{Semantic segmentation results of the vision foundation models and end-to-end trained model. We trained a UPerNet \cite{xiao2018unified} decoder on top of the frozen foundation model. We computed the mean and standard deviation of Dice similarity coefficients to evaluate segmentation performance. Values in \textbf{bold} indicate the best-performing results.}}
    \label{tab:seg}
    \begin{tabular}{c | c | c | c}
        \toprule
        \textbf{Models} & \textbf{Frozen Backbone} & \textbf{VinDr-RibCXR} & \textbf{Montgomery} \\ \midrule
        ConvNeXt \cite{liu2022convnet} & \ding{56} & 0.710 $\pm$ 0.116 & 0.954 $\pm$ 0.047 \\
        EVA-X \cite{yao2024eva} & \ding{52} & 0.471 $\pm$ 0.108 & 0.956 $\pm$ 0.036 \\
        RAD-DINO \cite{perez2024rad} & \ding{52} & 0.684 $\pm$ 0.094 & 0.978 $\pm$ 0.006 \\
        CheXFound & \ding{52} & \textbf{0.793 $\pm$ 0.062} & \textbf{0.980 $\pm$ 0.005} \\
        \bottomrule
    \end{tabular}
\end{table}

\begin{table}[t]
    \centering
    \setlength{\tabcolsep}{2pt}
    \setlength{\extrarowheight}{2pt}
    \caption{\blue{Benchmarking foundation models pretrained using different self-supervised methods on CXR-987K. We evaluated the downstream classification performance on CXR-LT 24. Values inside the parentheses are 95\% confidence intervals. Values in \textbf{bold} indicate the best-performing results.}}
    \label{tab:bmark}
    \begin{tabular}{c | c | c c}
        \toprule
        \textbf{Methods} & \textbf{Architectures} & \textbf{AUPRC} & \textbf{AUROC} \\ \midrule
        Medical MAE \cite{xiao2023delving} & ViT-B/16 & 0.193\tiny(.189--.197) & 0.772\tiny(.767--.779) \\
        iBOT \cite{zhou2021ibot}  & ViT-L/16 & 0.212\tiny(.207--.217) & 0.801\tiny(.796--.806) \\
        EVA-X \cite{yao2024eva}  & ViT-B/16 & 0.168\tiny(.165--.171) & 0.703\tiny(.698--.708) \\
        RAD-DINO \cite{perez2024rad} & ViT-B/14 & 0.185\tiny(.181--.189) & 0.751\tiny(0.743--.759) \\ \midrule
        CheXFound & ViT-L/16 & \textbf{0.265\tiny(.259--.271)} & \textbf{0.840\tiny(.836--.844)} \\
        \bottomrule
    \end{tabular}
\end{table}

\begin{table*}[t]
    \centering
    \caption{Model Performance across pretraining data sizes and model scales. Results are given in the mean values of AUPRC and AUROC over 1000 bootstrapped samples. Values inside the parentheses indicate the 95\% confidence intervals. Values in \textbf{bold} indicate the best-performing results.}
    \label{tab:scale}
    \setlength{\tabcolsep}{1.8pt}
    \setlength{\extrarowheight}{2pt}
    \begin{scriptsize}
    \begin{tabular}{c | c | c c | c c | c c | c c | c c}
    \toprule
    \textbf{Pretrain.} & \multirow{2}{*}{\textbf{Arch.}} & \multicolumn{2}{c}{\textbf{CXR-LT 24}} & \multicolumn{2}{c}{\textbf{CheXpert}} & \multicolumn{2}{c}{\textbf{Shenzhen}} & \multicolumn{2}{c}{\textbf{Montgomery}} & \multicolumn{2}{c}{\textbf{JSRT}} \\ \cmidrule(lr){3-12}
    \textbf{Data} & & \textbf{AUPRC} & \textbf{AUROC} & \textbf{AUPRC} & \textbf{AUROC} & \textbf{AUPRC} & \textbf{AUROC} & \textbf{AUPRC} & \textbf{AUROC} & \textbf{AUPRC} & \textbf{AUROC} \\ \midrule
    \multirow{2}{*}{CXR-207K} & ViT-Base & 0.185\tiny(.182–.188) & 0.775\tiny(.753--.789) & 0.598\tiny(.550–.645) & 0.854\tiny(.839--.862) & 0.921\tiny(.898--.946) & 0.909\tiny(.855–.953) & 0.889\tiny(.732--1.0) & 0.897\tiny(.742–1.0) & 0.756\tiny(.591–.882) & 0.612\tiny(.495--.712) \\ 
     & ViT-Large & 0.207\tiny(.203–.210) & 0.795\tiny(.774--.813) & 0.618\tiny(.629–.726) & 0.874\tiny(.862--.884) & 0.953\tiny(.923--.979) & 0.940\tiny(.901--.962) & 0.923\tiny(.811--1.0) & 0.932\tiny(.826--1.0) & 0.775\tiny(.620–.892) & 0.668\tiny(.537--.784) \\  \midrule
    \multirow{2}{*}{CXR-744K} & ViT-Base & 0.211\tiny(.208–.214) & 0.803\tiny(.792--.810) & 0.614\tiny(.571–.659) & 0.869\tiny(.854--.886) & 0.938\tiny(.913--.961) & 0.925\tiny(.875–.963) & 0.909\tiny(.775--1.0) & 0.915\tiny(.785–1.0) & 0.866\tiny(.759–.935) & 0.747\tiny(.695--.787) \\ 
    & ViT-Large & 0.217\tiny(.214–.221) & 0.813\tiny(.801--.822) & 0.643\tiny(.596–.692) & 0.887\tiny(.876--.896) & 0.964\tiny(.935–.986)) & 0.956\tiny(.923--.972) & 0.953\tiny(.843--1.0) & 0.957\tiny(.845--1.0) & 0.905\tiny(.794--.953) & 0.826\tiny(.674--.925) \\ \midrule
    \multirow{2}{*}{CXR-987K} & ViT-Base & 0.219\tiny(.215–.223) & 0.815\tiny(.811--.819)  & 0.632\tiny(.583–.679) & 0.877\tiny(.853--.896) & 0.957\tiny(.924--.979) & 0.947\tiny(.925–.963) & 0.923\tiny(.788–1.0) & 0.927\tiny(.818–1.0)  & 0.908\tiny(.806–.955) & 0.845\tiny(.696--.943) \\ 
    & ViT-Large & \textbf{0.265}\tiny(.259--.271) & \textbf{0.840}\tiny(.836--.844) & \textbf{0.679}\tiny(.630--.727) & \textbf{0.908}\tiny(.894--.921) & \textbf{0.983}\tiny(.960--.996) & \textbf{0.978}\tiny(.951--.995) & \textbf{1.000}\tiny(1.0--1.0) & \textbf{1.000}\tiny(1.0--1.0) & \textbf{0.986}\tiny(.956--1.0) & \textbf{0.975}\tiny(.931--1.0) \\
    \bottomrule
    \end{tabular}                
    \end{scriptsize}
\end{table*}

% \begin{itemize}
%     \item Evaluation of the generalizability of the model pretrained on 207K, 710K, and 971K CXRs on CXR-LT, CheXpert, Shenzhen, Montomery, and JSRT
%     \item Evaluation of the performance for the ViT-B (12 layers, 86M param.) and ViT-L (24 layers, 1024M param.) network architectures
% \end{itemize}

\begin{figure}[t]
    \centering
    \includegraphics[width=\linewidth]{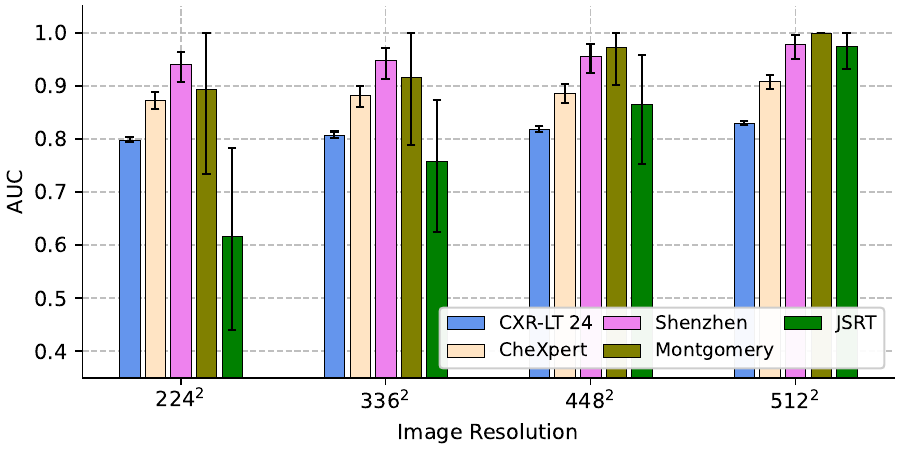}
    \caption{Evaluation of CheXFound model on the CXR-LT 24, CheXpert, Shenzhen, Montgomery, and JSRT datasets across a range of pretraining image resolutions. Error bars indicate 95\% confidence intervals.}
    \label{fig:role_res}
\end{figure}

\subsection{Benchmarking self-supervised pretraining methods}
% \blue{We benckmarked different self-supervised pretraining methods on our curated CXR-987K dataset to evaluate their effectiveness in representation learning. The use of a consistent CXR-987K dataset avoids the heterogeneity of data scales in the pretraining stage. We included various self-supervised methods for comparison, including Medical MAE \cite{xiao2023delving}, iBOT \cite{zhou2021ibot}, EVA-X \cite{yao2024eva}, and RAD-DINO \cite{perez2024rad}. 
% Classification performance was evaluated on the CXR-LT 24 dataset. As shown in Table IX below, CheXFound outperformed the next best-performing method (iBOT) by 3.9\% ($p < 0.001$) in AUROC. CheXFound achieved higher performance than iBOT owing to the additional advanced training approaches used by DINOv2, such as the KoLeo regularizer, Sinkhorn-Knopp centering, etc. Medical MAE underperformed CheXFound in part due to the absence of a discriminative objective that is critical to classification tasks. RAD-DINO, EVA-X, and Medical MAE are hindered by the ViT-Base architecture which is a shallower network architecture and encodes less effective representations compared with ViT-Large. 
% }
\blue{We benckmarked various self-supervised pretraining methods on our curated CXR-987K dataset to evaluate their effectiveness in representation learning. By consistently using the CXR-987K dataset, we eliminated the influence of data heterogeneity in the pretraining stage. The self-supervised methods for comparison include Medical MAE \cite{xiao2023delving}, iBOT \cite{zhou2021ibot}, EVA-X \cite{yao2024eva}, and RAD-DINO \cite{perez2024rad}. Classification performance was evaluated on the CXR-LT 24 dataset. As shown in Table \ref{tab:bmark} below, CheXFound outperformed the next best-performing method (iBOT) by 3.9\% ($p<0.001$) in AUROC. CheXFound achieved higher performance than iBOT owing to our advanced pretraining approach that utilized a low-resolution warm-up strategy and an optimized MIM loss weight on top of the original DINOv2. Medical MAE underperformed CheXFound because it lacks a discriminative objective which is critical for classification tasks. RAD-DINO, EVA-X, and Medical MAE were constrained by the shallower ViT-Base architecture, which encoded less effective representations than the ViT-Large architecture employed in our work.}

\subsection{Scalability of self-supervised vision encoders}
\label{sec:scale}
\black{The capabilities of the self-supervised vision encoders are affected by both the pretraining data scale and model scale \cite{oquab2023dinov2, chen2024towards}. To analyze the scaling trends, we pretrained CheXFound across a range of data scales, including CXR-987K and its two subsets CXR-744K and CXR-207K. We also evaluated the impact of the model scale by using ViT-Base (ViT-B) and ViT-Large (ViT-L) as the backbones.}
\blue{We maintained consistent hyperparameters across all pretraining data sizes and model capacities. We trained CheXFound for 100 epochs (2,500 iterations per epoch) using the AdamW optimizer with a batch size of 14 and an initial learning rate of 2e-4. The resolution of input chest X-rays was set to 512$\times$512. We also applied a low-resolution warm-up strategy and a masked image modeling loss weight of 3, as determined by the ablation studies in Section \ref{sec:hyerparam}.}

\black{Our results in Table~\ref{tab:scale} demonstrate that CheXFound benefits form both data and model scaling. Increasing the pretraining data from CXR-207K to CXR-987K with the ViT-L backbone leads to significant AUROC improvements of 3.5\% ($p<0.001$) on CXR-LT 24, 3.4\% ($p<0.001$) on CheXpert, 3.8\% ($p<0.001$) on Shenzhen, 6.8\% ($p<0.001$) on Montgomery, and 30.7\% ($p<0.001$) on JSRT. We observed similar trends when using ViT-B, which showed the performance consistently improved as we increased the pretraining data scale from CXR-207K to CXR-987K. In addition, CheXFound with ViT-L consistently outperformed the ViT-B architecture across different pretraining data scales. These results align with previous studies on scaling ViT models \cite{oquab2023dinov2, perez2024rad, chen2024towards}.}

\subsection{Impact of CXR resolution}

\black{%Increasing input CXR resolution is key to delivering improved performance in downstream tasks \cite{yang2024cardiovascular, perez2024rad}. 
To assess the impact of CXR resolution used for pretraining, we pretrained CheXFound using ViT-L with a patch size of 16 across a range of input resolution, including $224^2$, $336^2$, $448^2$, and $512^2$. 
% \py{These are image sizes but not resolutions. Convert to the resolution of images.} 
% \zf{Image resolution generally means image size in recent publications on self-supervised pretraining (such as DINOv2). I follow their terminology.}
We empirically found that pretraining at high resolution from scratch cannot produce meaningful representations for downstream tasks. To deal with this problem, we pretrained CheXFound at resolution $224^2$ from scratch and then used the pretrained weights to initialize higher-resolution pretraining at $336^2$, $448^2$, and $512^2$.}

\black{Fig. \ref{fig:role_res} shows that self-supervised pretraining at higher resolutions results in improved performance on downstream tasks. Increasing the pretraining resolutions from 224$^2$ pixels to 512$^2$ pixels significantly improves the AUROC by 3.3\% ($p<0.001$), 3.5\% ($p<0.001$), 3.7\% ($p<0.001$), 10.7\% ($p<0.001$), and 35.8\% ($p<0.001$) on the CXR-LT 24, CheXpert, Shenzhen, Montgomery, and JSRT datasets, respectively.}
% \py{on what???}
% \begin{itemize}
%     \item Evaluate performance of the self-supervised vision encoders (ViT-L) at 224, 336, 448, 512 on CXR-LT
% \end{itemize}

% \subsection{Role of feature combination}
% \begin{itemize}
%     \item Ablation studies on how many blocks of features used for final predictions
% \end{itemize}
\subsection{Ablation studies on pretraining}
\label{sec:hyerparam}
\begin{figure}[t]
    \centering
    \includegraphics[width=\linewidth]{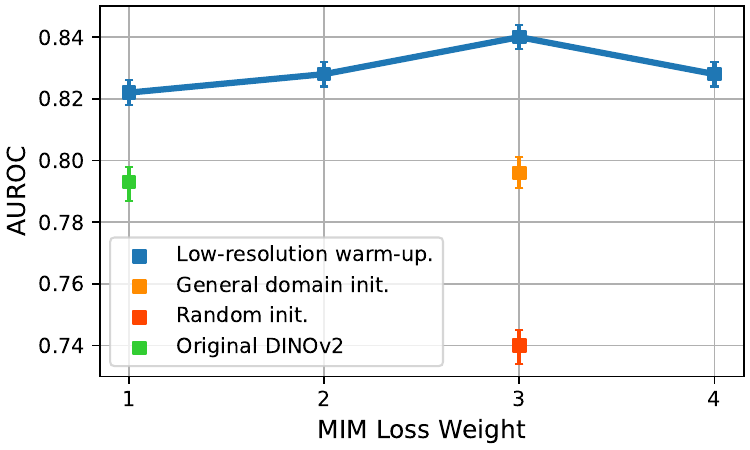}
    \caption{\blue{Ablation studies on the impact of MIM loss weights and parameter initialization strategies on downstream classification. We evaluated the classification performance on the CXR-LT 24 dataset. Error bars indicate 95\% confidence intervals.}}
    \label{fig:ibot_weight}
\end{figure}

% \blue{
% % We modified the original DINOv2 training strategies to make the model learn more effective representations for downstream chest X-ray analysis tasks.
% We evaluated both the impact of different parameter initialization strategies and MIM loss weights. As shown in Fig. \ref{fig:ibot_weight}, pretraining from scratch and general domain initialization that uses the checkpoint pretrained on natural images achieved AUROCs of 0.740 and 0.796 for classifying 40 diseases on CXR-LT 24 respectively. Low-resolution initialization that initializes pretraining at 512$^2$ resolution with the checkpoint pretrained at 224$^2$ resolution achieved an AUROC of 0.840 which outperformed pretraining from scratch and general domain initialization by large margins.
% We conjecture that the performance of high-resolution pretraining was hindered by the difficulty to summarize extensive contextual information and this issue was mitigated by warming up the model with the low-resolution model that captures high-level contextual information. Fig. \ref{fig:ibot_weight} also shows that increasing MIM loss weights enhances the performance in AUROC from 0.822 to 0.840, which demonstrates the effectiveness of encouraging patch tokens to encode local features of chest X-rays.
% }
\blue{We evaluated the impact of various parameter initialization strategies on downstream performance. As shown in Fig. \ref{fig:ibot_weight}, random initialization and general domain initialization that utilized a checkpoint pretrained on natural images achieved AUROCs of 0.740 and 0.796, respectively, for classifying 40 diseases on CXR-LT 24. Incorporating a low-resolution warm-up strategy, which initializes pretraining at 512$\times$512 resolution with the checkpoint pretrained at 224$\times$224 resolution, increased the AUROC to 0.840, outperforming both the random initialization and general domain initialization strategies by large margins. We hypothesize that the degraded performance of direct high-resolution pretraining was due to the limited capability of the ViT architecture to capture extensive contextual information, which was mitigated by warming up the model with the low-resolution pretrained model that had already learned high-level contextual representations.}

\blue{We further evaluated the impact of pretraining loss weights. The original DINOv2 settings employ balanced weights of 1:1 for the [CLS] token alignment and MIM losses for representation learning with natural images, where local patches often contain sufficient information to infer the global view. In contrast, medical images are characterized by subtle differences, with diagnostic cues often residing in nuanced local features. As such, we conducted multiple runs of pretraining with increased MIM loss weights to search for an optimal weight factor that adequately encourages local feature learning. As illustrated in Fig. \ref{fig:ibot_weight}, increasing MIM loss weights from 1 to 3 improved the AUROC from 0.822 to 0.840.}

\blue{Overall, compared with the original DINOv2 settings that achieved an AUROC of 0.793 (Fig. \ref{fig:ibot_weight}) by training the model from scratch at 224$\times$224 resolution with a MIM loss weight of 1, our modifications to the initialization strategy for high-resolution pretraining and MIM loss weight delivered a performance gain of 4.7\%.}

\begin{table}[t]
    \centering
    \setlength{\tabcolsep}{2pt}
    \setlength{\extrarowheight}{2pt}
    \caption{\blue{Ablation studies on the GLoRI architecture. We evaluated the classification performance on CXR-LT 24. We computed the mean AUPRC and AUROC values and their 95\% confidence intervals over 1,000 bootstrapped samples. Values in \textbf{bold} indicate the best-performing results.}}
    \label{tab:abl_glori}
    \begin{tabular}{l | c c c c}
        \toprule
        \multicolumn{1}{c}{\textbf{Architecture}} & \textbf{AUPRC} & $p$-value & \textbf{AUROC} & $p$-value \\ \midrule
        Linear probe & 0.209\tiny(.204–.214) & - & 0.799\tiny(.794--.804) & - \\
        \midrule
        Attention pooler & 0.246\tiny(.241--.252) & - & 0.826\tiny(.822--.830) & - \\
        + Global representation & 0.252\tiny(.247--.258) & 0.023 & 0.830\tiny(.826--.834) & 0.021 \\
        + Adaptive Temperatures & 0.262\tiny(.256--.268) & \footnotesize{$<0.001$} & 0.835\tiny(.831--.839) & \footnotesize{$<0.001$} \\
        + Pyramid Patch Merging & \textbf{0.265}\tiny(.259--.271) & \footnotesize{$<0.001$} & \textbf{0.840}\tiny(.836--.844) & \footnotesize{$<0.001$} \\
        \bottomrule
    \end{tabular}
\end{table}

\begin{figure*}[t]
    \centering
    \includegraphics[width=\linewidth]{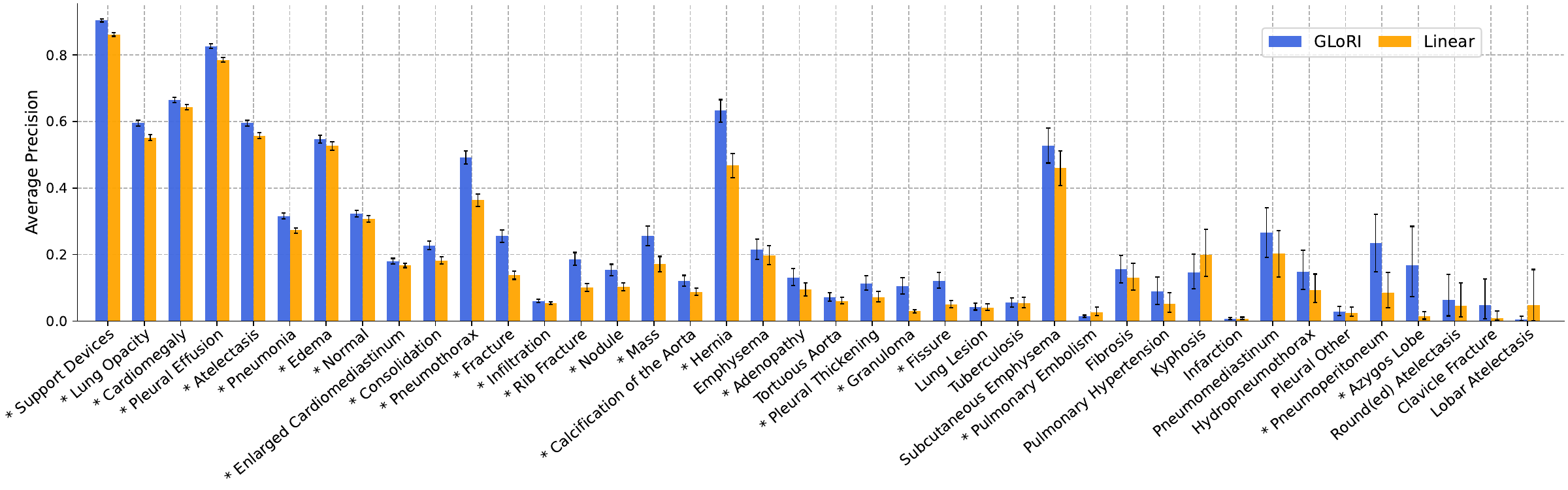}
    \caption{\blue{Detailed results comparing GLoRI and linear probing for classifying different diseases. Diseases are sorted in descending order regarding their frequency on the dataset. We computed the mean and 95\% confidence intervals of average precision using 1,000 bootstrapped samples. Error bars indicate 95\% confidence intervals. Asterisks indicate statistical significance ($p<0.05$).”}}
    \label{fig:role_glori}
\end{figure*}

\begin{figure*}[t]
    \centering
    \includegraphics[width=\textwidth]{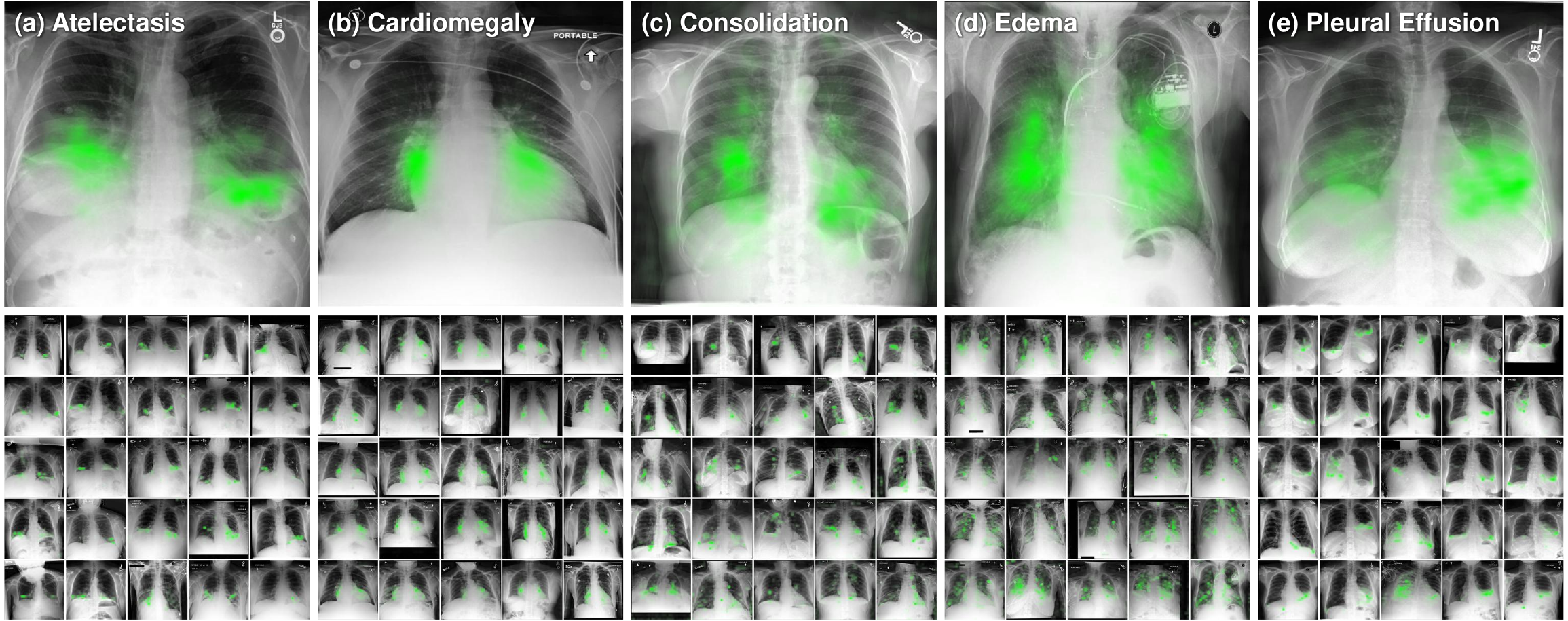}
    \caption{GLoRI attention maps for disease findings of atelectasis, cardiomegaly, consolidation, edema, and pleural effusion, respectively. Each subfigure contains 25 CXRs with their GLoRI attention maps overlaid (bottom) and an anchor CXR with a global attention map overlaid (top).
    % \red{randomly selected [correct? please confirm.]}
    %Green regions in the attention maps indicate critical anatomical locations. The 25 CXRs are with a single disease finding and high predictive probabilities from our CheXFound model. 
    The global attention maps are the averaged attention maps of 25 CXRs after registered to the anchor image. 
    % \py{please reduce the caption font size.}
    % \py{Be consistent on using CXR or chest X-ray. Use one or the other, but not both.}
    }
    \label{fig:glori-attns}
\end{figure*}

\subsection{Ablation studies on GLoRI head}
\blue{We proposed GLoRI, a prediction head that incorporates both global and local image features for multilabel disease classification. Compared with prior attention pooling approaches \cite{moutakanni2024advancing, bardes2024revisiting} that rely solely on an attention pooler with multi-head attention, GLoRI enhances classification performance by integrating an adaptive temperature module and a pyramid patch merging module into the multi-head attention mechanism and by incorporating the global image features. We evaluated GLoRI on the CXR-LT 24 dataset. As shown in Table \ref{tab:abl_glori}, the attention pooler with a single multi-head attention layer achieved an AUROC of 0.826. This performance was improved by 0.4\%, 0.5\%, and 0.5\% by incrementally incorporating the global representation, adaptive temperatures, and pyramid patch merging, respectively. Eventually, GLoRI achieved an AUROC of 0.84, which significantly improved the attention pooler by 1.4\% ($p < 0.001$, two-sided permutation test) and outperformed linear probe by 4.1\% ($p < 0.001$). We compared GLoRI with linear probe on detecting 40 diseases in Fig. \ref{fig:role_glori}. Overall, these improvements are attributed to GLoRI’s enriched representations, which integrate both the fine-grained and coarse-grained local features alongside the global image features.}

\section{Discussion and Conclusion}
\subsection{Interpretation of disease-specific local features}
\label{sec:intepret}
\black{The interpretability of an artificial intelligence model is crucial to its medical applications. 
In this study, we trained GLoRI on top of the frozen foundation model. GLoRI inherently provides interpretable attention maps for each pathology. We visualized the attention maps for a selection of five pathologies of 25 CXRs and aligned these attention maps to an anchor CXR via affine registration\footnote{We apply affine registration using the SimpleElastix library: \url{https://simpleelastix.readthedocs.io}.} to provide a global perspective in Fig. \ref{fig:glori-attns}. The attention maps contain precise localization of abnormalities in CXRs and the global attention maps cover the regions where the pathologies constantly occur. 
To be specific, edema refers to the accumulation of excess fluid and its abnormal regions often diffuse across lungs. This pattern is well captured by our attention maps as we observed in Fig. \ref{fig:glori-attns} (d) that the critical regions in the individual attention maps scatter over the lungs and the global attention map covers extensive regions of both lungs.
However, these attention maps have the limitations of covering partial regions of the abnormalities. For example, the attention maps for cardiomegaly only cover the heart on the left and right regions of the spine, and some maps for pleural effusion only cover the inferior boundaries of the lung while ignore the remaining abnormal regions.}

% \py{affine or rigid? What tool did you use?}
% limitations
%Overall, we demonstrate that with rich representations learned through large-scale self-supervised pretraining, deep learning models show improved interpretability.

% \begin{itemize}
%     \item significance of interpretable artificial intelligence models, trustworthiness
%     \item CAM, GradCAM, weighted average of feature maps, lack localization precision
%     \item GLoRI with query tokens, can provide attention maps for the attribution of model prediction.
%     \item limitations, attention regions
% \end{itemize}

\subsection{Generalizability of foundation models}
\black{
An important characteristic of CheXFound and other foundation models is their generalization capabilities to in-distribution and out-of-distribution downstream tasks. Compared with other encoders, we found that CheXFound achieved better performances on both in-distribution CXR-LT 24 and CheXpert datasets and out-of-distribution Shenzhen, Montgomery, and JSRT datasets. On the opportunistic CXR interpretation tasks on PLCO, CheXFound also achieved consistent and significant increases over comparison methods. CheXFound's generalizability is attributed to the strong representation quality of frozen features learned via pretraining with large-scale, diverse CXRs. We also demonstrated CheXFound's generalization capabilities on infrequent and underrepresented pathologies. CheXFound achieved significant increases over its comparisons in classifying low-prevalence pathologies with lower than 1\% occurrence frequencies, demonstrating its superior label efficiency. Although CheXFound with ViT-L achieved robust generalizability, our study did not evaluate the best-performing ViT-giant (ViT-g) architecture in DINOv2, a larger model with 1.1B parameters, which we expect to achieve better generalization performances in CXR interpretation, but it demands more pretraining data and computational resources. Overall, we demonstrated CheXFound's robust generalization capabilities, which we believe can enable diverse downstream adaptations with improved label efficiency.}

\subsection{Conclusion}
\black{In summary, this work introduces CheXFound, a vision-centric foundation model pretrained via self-distillation on over a large cohort of unique CXRs. For downstream adaptations, we trained a GLoRI head on top of the frozen CheXFound, which combined fine- and coarse-grained disease-specific local features and global image features to improve the multilabel classification performance. CheXFound outperformed previous methods for classifying 40 disease findings on CXR-LT 24. CheXFound demonstrated superior label efficiency on datasets with limited training labels and strong generalization capabilities to adapt to opportunistic CXR interpretation, malposition tube detection, and anatomical structure segmentation.
The disease-specific local features extracted from CheXFound also manifested strong interpretability as characterized by the attention maps.
In our future work, we will continue to explore novel pretraining schemes to further improve the understanding of CXRs by the foundation models.}
{\color{black}
\bibliographystyle{IEEEtran}
\bibliography{ref}
}
\end{document}